\newcommand{\eqnstart}{\begin{equation}}
\newcommand{\eqnend}{\end{equation}}
\documentclass[twocolumn,tighten]{aastex61}
\usepackage{float}
\usepackage{amsmath}
\usepackage{color}
\usepackage[normalem]{ulem}

\usepackage{bm}
\usepackage[utf8]{inputenc}

\newcommand{\editadded}[1]{{\color{red}\textbf{#1}}}
\newcommand{\editremoved}[1]{\sout{#1}}
\renewcommand{\editadded}[1]{#1}
\renewcommand{\editremoved}[1]{}

\newlength{\figurewidth}
\newlength{\wfigurewidth}
\DeclareMathOperator{\sign}{sign}

\newcommand{\figstart}{\begin{figure}\begin{center}}
\newcommand{\figend}{\end{center}\end{figure}}
\newcommand{\wfigstart}{\begin{figure*}\begin{center}}
\newcommand{\wfigend}{\end{center}\end{figure*}}

\begin{document}

\title{The HAWC Real-Time Flare Monitor for Rapid Detection of Transient Events}
\shorttitle{HAWC Real-Time Flare Monitor}

\correspondingauthor{T. Weisgarber}
\email{weisgarber@wisc.edu}
\AuthorCollaborationLimit=300

\author{A.U.~Abeysekara}\affil{Department of Physics and Astronomy, University of Utah, Salt Lake City, UT, USA}
\author{R.~Alfaro}\affil{Instituto de F{\'i}sica, Universidad Nacional Aut{\'o}noma de M{\'e}xico, Mexico City, Mexico}
\author{C.~Alvarez}\affil{Universidad Aut{\'o}noma de Chiapas, Tuxtla Guti{\'e}rrez, Chiapas, Mexico}
\author{J.D.~{\'A}lvarez}\affil{Universidad Michoacana de San Nicol{\'a}s de Hidalgo, Morelia, Mexico}
\author{R.~Arceo}\affil{Universidad Aut{\'o}noma de Chiapas, Tuxtla Guti{\'e}rrez, Chiapas, Mexico}
\author{J.C.~Arteaga-Vel{\'a}zquez}\affil{Universidad Michoacana de San Nicol{\'a}s de Hidalgo, Morelia, Mexico}
\author{D.~Avila Rojas}\affil{Instituto de F{\'i}sica, Universidad Nacional Aut{\'o}noma de M{\'e}xico, Mexico City, Mexico}
\author{H.A.~Ayala Solares}\affil{Department of Physics, Michigan Technological University, Houghton, MI, USA}
\author{A.S.~Barber}\affil{Department of Physics and Astronomy, University of Utah, Salt Lake City, UT, USA}
\author{N.~Bautista-Elivar}\affil{Universidad Politecnica de Pachuca, Pachuca, Hidalgo, Mexico}
% ...

\begin{abstract}

We present the development of a real-time flare monitor for the High Altitude Water Cherenkov (HAWC) observatory.
The flare monitor has been fully operational since 2017 January and is designed to detect very high energy (VHE; $E\gtrsim100$ GeV) transient events from blazars on time scales lasting from 2 minutes to 10 hours in order to facilitate multiwavelength and multimessenger studies.
These flares provide information for investigations into the mechanisms that power the blazars' relativistic jets and accelerate particles within them, and they may also serve as probes of the populations of particles and fields in intergalactic space.
To date, the detection of blazar flares in the VHE range has relied primarily on pointed observations by imaging atmospheric Cherenkov telescopes.
The recently completed HAWC observatory offers the opportunity to study VHE flares in survey mode, scanning 2/3 of the entire sky every day with a field of view of $\sim$1.8 steradians.
In this work, we report on the sensitivity of the HAWC real-time flare monitor and demonstrate its capabilities via the detection of three high-confidence VHE events in the blazars Markarian 421 and Markarian 501.

\end{abstract}
\keywords{astroparticle physics --- gamma rays: galaxies --- galaxies: BL Lacertae objects: individual (Mrk 421, Mrk 501) --- methods: data analysis}

\section{Blazar Transients at Very High Energies}\label{sec:transients}

Active galactic nuclei (AGNs) comprise a class of extremely luminous galactic cores that are widely understood to be powered by accretion onto a supermassive black hole \citep{Begelman:1984fm}.
Many AGNs exhibit highly collimated jets, first observed by \citet{Curtis:1918vg} in the object M87, that efficiently transport plasma via bulk relativistic outflows \citep{Boettcher:2012tv}.
In a small number of AGNs, the jet is serendipitously aligned along the line of sight to Earth.
Under the standard AGN unification scheme \citep{Urry:1995ky}, such objects are referred to as blazars, and their enhanced emission due to Lorentz boosting serves as a probe into the production and dynamics of relativistic jets.

Perhaps the most striking feature of blazars is their extreme variability.
Long-term observations in the VHE band reveal that the flux may reach levels several times higher than the quiescent emission and persist in such states for years \citep{Tluczykont:2007bt,Acciari:2014db}.
Extreme VHE flares surpassing the quiescent emission by factors of 100 and with observed variability time scales as short as minutes have also been observed in some sources \citep[e.g.][]{Aharonian:2007ep}.
These flares are of particular interest because they facilitate studies of the mechanisms powering the blazar central engine \citep{Katarzynski:2005ek,Madejski:2016ig} and they place stringent constraints on the bulk jet Lorentz factors \citep{Spada:2001es,Krawczynski:2001gf,Begelman:2008cq}.
They also constrain the location of the region where electromagnetic energy in the initially Poynting-dominated jet flux~\citep{Blandford:1977ir,Blandford:1982ca} is dissipated into accelerated particles, producing flares~\citep{Nalewajko:2014fh}.

In general, multiwavelength observations providing spectral and, where possible, morphological or polarization information during and after a flare have yielded the most complete pictures of the dissipation process \citep[e.g.][]{Marscher:2008ii,Acciari:2009hv,Abdo:2010gx}.
Developing an understanding of dissipation in jets will in turn help explain jet composition and structure, accretion, jet formation, and AGN feedback \citep{Bykov:2012bu}.
Furthermore, the increased fluxes and short variability time scales defining extreme VHE flares render them useful for studying Lorentz invariance violation \citep{Biller:1999dm,Albert:2008fd,Abramowski:2011hl}, the extragalactic background light \citep[EBL; e.g.][]{Mazin:2007dh,Biteau:2015kz}, or the intergalactic magnetic field \citep{Aharonian:1994ix,Plaga:1995hv,Neronov:2009fo}.

Although many informative studies on extreme VHE flares have been performed, until the present time no dedicated VHE survey instrument has been available for monitoring large numbers of blazars simultaneously.
Imaging atmospheric Cherenkov telescopes (IACTs) are the most sensitive VHE instruments, but their narrow fields of view prohibit unbiased and continuous monitoring of more than a few sources.
The Large Area Telescope (LAT) on board the Fermi Gamma-Ray Space Telescope (hereafter Fermi-LAT) surveys the sky in the energy range from 100 MeV to 300 GeV \citep{Atwood:2009ha}, but flares detected by the Fermi-LAT are not known {\it a priori} also to produce VHE emission.
Consequently, the advent of the High Altitude Water Cherenkov (HAWC) observatory, the most sensitive VHE survey instrument yet constructed, opens a new window of opportunity for identifying extreme VHE flares.
The rapid identification of these events with HAWC permits follow-up observations with the more sensitive IACTs---VERITAS, HESS, MAGIC, and FACT---as well as with instruments sensitive to other energies or other particle types, such as Swift, Fermi, NuSTAR, Chandra, and IceCube.

Many VHE flares exhibit soft spectra with photon indices $\Gamma$ in the range from 3--4 \citep[e.g.][]{Arlen:2013iw,Abeysekara:2017bl} for power-law spectra of the form $dN/dE\propto E^{-\Gamma}$.
However, some flares have been observed with harder indices, in the range from 2--3, either over the entire VHE range \citep{Archambault:2015eq,2016ATel.9010....1B} or as a hard low-energy component up to several hundred GeV \citep{Aharonian:2007ep}.
This makes them good candidates for detection by a dedicated survey instrument such as HAWC.
Since the most luminous flares (e.g. those that exceed the flux from the Crab Nebula) tend to have short durations, obtaining a multiwavelength picture of the brightest transients requires a rapid alert system capable of issuing alerts before the transient activity has finished.

In this work, we present the development and deployment of a real-time flare monitor designed to identify extreme VHE flares with the HAWC observatory and issue alerts rapidly to other instruments.
Following a brief summary of HAWC operations in Section~\ref{sec:hawc-operation}, we present the mechanics of the flare monitor in Section~\ref{sec:method}, describe the tuning of the false alarm rate in Section~\ref{sec:tuning}, and characterize the flare monitor sensitivity in Section~\ref{sec:sensitivity}.
Section~\ref{sec:targets} describes the selection of flare monitor targets.
In Section~\ref{sec:verify}, we apply the flare monitor to archival data on the two nearest blazars, Markarian 421 and Markarian 501 (hereafter Mrk 421 and Mrk 501), located respectively at redshifts of $z=0.031$ and $z=0.033$ \citep{Mao:2011ig}.
The flare monitor identifies three high-confidence events in these data.
We discuss our future plans for the flare monitor in Section~\ref{sec:outlook}.

\section{The HAWC Observatory}\label{sec:hawc-operation}

The HAWC observatory consists of a close-packed array of 300 optically isolated water Cherenkov detectors, situated at 4100 m above sea level in the state of Puebla, Mexico.
Located at $19^\circ$ north latitude, HAWC surveys 2/3 of the entire sky with an uptime in excess of $90\%$, making it an ideal instrument for searching for VHE transient phenomena.
Each water Cherenkov detector contains four upward-facing photomultiplier tubes (PMTs) that sample the energy deposition at ground level from air showers initiated by the interactions of primary particles with the atmosphere.
HAWC detects events in the energy range from $\sim$100 GeV to $>$100 TeV.
Further details concerning the performance of HAWC may be found in \citet{HAWC-Crab-paper}.

Over $99.9\%$ of the events that HAWC detects are cosmic rays that must be separated efficiently from any gamma-ray signal present in the data.
Due to the large transverse momentum carried by the products of hadronic interactions in the air showers, the distribution of energy is much less smooth for showers initiated by cosmic rays than it is for gamma-ray showers.
To remove a large fraction of the cosmic rays, we apply gamma-hadron separation cuts in two parameters designed to characterize the smoothness of the energy distribution.
As described in \cite{HAWC-Crab-paper}, one parameter tests for a single localized excess far from the shower core, while another tests for rotational symmetry about the shower axis.

We apply the gamma-hadron separation cuts separately to analysis bins defined by the fractional number of operational PMTs that participate in the event.
The present work includes the nine analysis bins used in \cite{HAWC-Crab-paper} but also includes events that trigger between $4.4\%$ and $6.7\%$ of the detector in an additional bin.
The bins are numbered from bin 0, in which the smallest fraction of PMTs participate, to bin 9, in which nearly every PMT in the detector participates.
\editadded{Table~\ref{sec:hawc-operation:tab:analysis-bins} shows the high and low fractions of operational PMTs that define each of the ten bins.}

\begin{deluxetable}{cccc}
\tablecaption{\editadded{Definition of Analysis Bins}\label{sec:hawc-operation:tab:analysis-bins}}
\tablehead{
\colhead{} & \colhead{\editadded{Low}} & \colhead{\editadded{High}} & \colhead{\editadded{Smoothing}} \\
\colhead{\editadded{Bin}} & \colhead{\editadded{Fraction}} & \colhead{\editadded{Fraction}} & \colhead{\editadded{Radius [$^\circ$]}}
}
\startdata
\editadded{0} & \editadded{0.044} & \editadded{0.067} & \editadded{1.20} \\
\editadded{1} & \editadded{0.067} & \editadded{0.105} & \editadded{1.20} \\
\editadded{2} & \editadded{0.105} & \editadded{0.162} & \editadded{0.75} \\
\editadded{3} & \editadded{0.162} & \editadded{0.247} & \editadded{0.60} \\
\editadded{4} & \editadded{0.247} & \editadded{0.356} & \editadded{0.40} \\
\editadded{5} & \editadded{0.356} & \editadded{0.485} & \editadded{0.40} \\
\editadded{6} & \editadded{0.485} & \editadded{0.618} & \editadded{0.30} \\
\editadded{7} & \editadded{0.618} & \editadded{0.740} & \editadded{0.30} \\
\editadded{8} & \editadded{0.740} & \editadded{0.840} & \editadded{0.30} \\
\editadded{9} & \editadded{0.840} & \editadded{1} & \editadded{0.25} \\
\enddata
\tablecomments{\editadded{Low and high fraction refer to the fraction of active PMTs participating in the event. The smoothing radius describes a top-hat function used to smooth both the on-source and off-source events.}}
\end{deluxetable}

\editadded{
Each event that triggers at least $4.4\%$ of the operational PMTs falls into exactly one analysis bin.
When determining the events that are associated with a given source, we smooth the events with a top-hat function with a radius that depends on the analysis bin.
The radius that maximizes the significance of the Crab Nebula, the strongest steady source of VHE gamma rays, appears in the fourth column of Table~\ref{sec:hawc-operation:tab:analysis-bins}.
}

While other recent HAWC publications \citep[e.g.][]{HAWC-Crab-paper,HAWC-catalog-paper} report results in terms of source fluxes, the real-time flare monitor uses only the observed event counts, not fluxes, to decide whether to issue an alert.
It is therefore less affected by an existing discrepancy between observed event counts and Monte Carlo simulation in bin 0 that prevents the inclusion of bin 0 events in the flux estimation.
Efforts are currently ongoing to resolve this discrepancy.
Bin 0 generally contains the lowest energy gamma rays that are of considerable interest to the study of blazars.
We have tested the effect of including bin 0 by performing simulations as described in Section~\ref{sec:sensitivity} with and without the bin 0 data included.
We find that the inclusion or exclusion of bin 0 does not affect the sensitivity of the real-time flare monitor, and therefore we include it in the present analysis.

In general, events with larger energies and smaller zenith angles tend to fall into the higher analysis bins.
However, there is substantial overlap in the distribution of energies within the bins, so the analysis bins cannot be interpreted as strict energy bins.
Instead, the analysis bins correspond more directly to the amount of information available about a given event, with the higher analysis bins providing more information and better angular resolution.
This renders the analysis bins suitable for the likelihood method described in the following section.
\editadded{As was done for the smoothing radius,}\editremoved{The values of} \editadded{we tune} the cuts in each analysis bin \editremoved{are tuned} to maximize the sensitivity to the Crab Nebula\editadded{.}\editremoved{, the strongest steady source of VHE gamma rays.}

While many AGNs exhibit spectra harder than that of the Crab at lower energies, these are softened at higher energies due to EBL absorption.
Typical EBL models \citep[e.g.][]{Franceschini:2008em,Dominguez:2010do,Gilmore:2012il} indicate that the attenuation of gamma rays via pair production becomes of order unity around a redshift of $z\approx1$ for an observed energy of 100 GeV, or a redshift of $z\approx0.2$ for gamma rays with energies of $\sim$400 GeV.
Furthermore, in contrast to that of the IACTs, HAWC's sensitivity varies slowly with energy, reaching its maximum only in the multi-TeV regime \citep[see Figure 14 of][]{HAWC-Crab-paper}.
As a result of these effects, HAWC is most sensitive to hard flares from nearby blazars.
Since several known VHE blazars lie at redshifts $z\lesssim0.3$, and intrinsically hard ($\Gamma\lesssim2$) VHE spectra are likely to be rendered softer than the Crab, we expect that the sensitivity reported here is somewhat optimistic for the detection of realistic blazar flares.
At present, we reserve an in-depth investigation into the precise sensitivity to different flare properties for future work.

\section{Flare Monitor Search Method}\label{sec:method}

The HAWC real-time flare monitor is a fully automated system capable of rapid transient detection on minute to hour time scales.
It complements other ongoing HAWC search efforts that are tuned to longer time scales or focused on gamma-ray bursts.
As described in Section~\ref{sec:targets}, we choose a number of targets to monitor based on their high probability to produce VHE gamma rays.
We collect the data from each monitored target into a buffer containing 300 observations.
Each observation has a duration of 2 minutes, which is consequently the minimum time resolution for the method.
The buffer depth of 300 observations is chosen as a compromise between computer memory utilization and providing a sufficient number of observations to detect flares lasting up to a few hours.
We record data in the buffer exclusively at times for which the target is located at zenith angles less than $45^\circ$.
As a result, the 10 hours of buffer time correspond to a much larger amount of real time that includes at least 4 hours of time from the previous day's observations.
For targets with declinations far from the latitude of HAWC, the buffer contains several days of observations.

Following the language of \citet{Li:1983de}, we use an off-source region to estimate the background contribution to the event count in the on-source region.
The off-source count is derived from the direct integration technique \citep{Atkins:2003kw} with an integration time of 2 hours.
Because the flare monitor must operate in real time, the result of the direct integration from the previous 2 hours is used when deriving the background estimation for the current observations.
Most of the time, the response of the detector to air showers is sufficiently stable that the distribution of reconstructed events in horizontal (Earth-fixed) coordinates does not change significantly over the course of 4 hours.
This permits us to use the previous direct integration result.
We address occasional changes in the detector response in Section~\ref{sec:tuning:sub:quality}.

\subsection{Likelihood Ratio Test for Flare Detection}\label{sec:method:sub:lrt}

Each time the buffer is updated with a new observation, we perform a likelihood ratio test to search for an increase in the observed flux.
In this test, the null hypothesis is that the data can be represented by a constant flux for the entire duration of the buffer.
For the alternative hypothesis, we assume that the flux is different at different points in the buffer.
In what follows, we first derive the likelihood ratio for the case when the buffer is divided into any number of distinct flux states.
We then specialize to the case for which there are only two states: an initial low state followed by a high state.
This is the case of interest for the detection of a flare.

We begin by considering a single analysis bin $b$.
For time bin $i$, we represent the observed number of on-source events in the analysis bin as $n_i$ and the observed number of off-source events as $m_i$.
We assume that the on-source count $n_i$ is derived from a Poisson distribution with mean $\nu_i$, and that the off-source count $m_i$ is derived from a Poisson distribution with mean $\mu_i$.
Following \cite{Li:1983de}, we designate the ratio of on-source to off-source exposure as $\alpha$.
The value of $\alpha$ depends on the analysis bin, ranging from $0.082$ for bin 0 down to $0.017$ for bin 9.
The probability to observe the measured counts given the model parameters for time bins running from $j$ to $k$ inclusive is then
\eqnstart\label{sec:method:eqn:base-likelihood}
P(\bm{n},\bm{m}|\bm{\nu},\bm{\mu})=\prod_{i=j}^k\frac{e^{-\nu_i}\nu_i^{n_i}}{n_i!}\frac{e^{-\mu_i}\mu_i^{m_i}}{m_i!}=L_{b;j,k},
\eqnend
for analysis bin $b$.
Here, $L_{b;j,k}$ is the likelihood of the model specified by $\bm{\nu}$ and $\bm{\mu}$ giving rise to the observations in the specified range of time bins.
For clarity, we omit the subscript $b$ on $n_i$, $m_i$, $\nu_i$, $\mu_i$, and $\alpha$.

Let us now assume that we have $T$ time bins in the buffer, and that the observations are partitioned into $B$ distinct flux states, with $1\leq B\leq T$.
The null hypothesis is thus formulated as $B=1$ and the alternative hypothesis as $B>1$.
Let the position of the lower edge of flux state $j$ in the buffer be denoted by $l_j$ and the upper edge by $u_j$, such that the flux state runs from time bins $l_j$ to $u_j$, inclusive.
Clearly, $l_1=1$ and $u_B=T$, and since the flux states are not overlapping, $l_{j+1}=1+u_j$.
The total likelihood of the model for bin $b$ is then the product of the likelihoods for each flux state:
\eqnstart\label{sec:method:eqn:total-likelihood}
L^{\mathrm{tot}}_b=\prod_{i=1}^BL_{b;l_i,u_i}.
\eqnend
Although we have assumed a constant flux in each state, Equation~\ref{sec:method:eqn:total-likelihood} can accommodate any flare shape with a time resolution equal to the time bin size of 2 minutes.

It is convenient to re-cast Equation~\ref{sec:method:eqn:base-likelihood} in terms of the ratio $\rho_i\equiv\nu_i/(\alpha\mu_i)$.
In this case, the log-likelihood takes the form
\eqnstart\label{sec:method:eqn:base-log-likelihood}
\begin{split}
\ln L_{b;j,k}=&\sum_{i=j}^k\left[-\alpha\mu_i\rho_i+n_i\ln(\alpha\mu_i\rho_i)-\mu_i\right.\\
&\left.+m_i\ln\mu_i-S_i\right],
\end{split}
\eqnend
where $S_i=\ln(n_i!)+\ln(m_i!)$.

Adopting a frequentist approach, we maximize the likelihood with respect to the parameters $\mu_i$ and $\rho_i$.
If we assume that the flux is constant from time bin $j$ through time bin $k$, then $\rho_i$ takes on a single, constant value $\rho_{j,k}$.
Since $\rho_i$ is constructed as a ratio of on-source to off-source counts, the dependence of the detector response on the zenith angle as the target moves across the sky is canceled.
We allow the values of $\mu_i$ to vary from time bin to time bin because we expect these to be affected by the detector response, which depends on the instantaneous zenith angle of the target.
The solutions that maximize the likelihood are
\eqnstart\label{sec:method:eqn:maximize-likelihood-solution}
\begin{aligned}
\rho_{j,k}=\frac{N_{j,k}}{\alpha M_{j,k}}; && \mu_i=\frac{n_i+m_i}{N_{j,k}+M_{j,k}}M_{j,k},
\end{aligned}
\end{equation}
where
\begin{equation}\label{sec:method:eqn:sum-definition}
\begin{aligned}
N_{j,k}\equiv\sum_{i=j}^kn_i; && M_{j,k}\equiv\sum_{i=j}^km_i.
\end{aligned}
\eqnend
With these solutions, the expression for the maximum log-likelihood becomes
\eqnstart\label{sec:method:eqn:maximize-likelihood}
\begin{split}
\ln L^{\mathrm{max}}_{b;j,k}=&-N_{j,k}-N_{j,k}\ln\left(1+\frac{M_{j,k}}{N_{j,k}}\right)\\
&-M_{j,k}-M_{j,k}\ln\left(1+\frac{N_{j,k}}{M_{j,k}}\right)\\
&+\sum_{i=j}^{k}\left[(n_i+m_i)\ln(n_i+m_i)-S_i\right].
\end{split}
\eqnend

The log of the ratio of the alternative hypothesis likelihood to the null hypothesis likelihood is given by
\eqnstart\label{sec:method:eqn:lrt}
D_b=\Delta\ln L^{\mathrm{max}}_b=\sum_{i=1}^{B}\ln L^{\mathrm{max}}_{b;l_i,u_i}-\ln L^{\mathrm{max}}_{b;1,T}.
\eqnend
Here, $D_b$ is a test statistic that is equal to half the usual test statistic for the likelihood ratio test, $2\Delta\ln L_b$.
By inspection of Equation~\ref{sec:method:eqn:maximize-likelihood}, we see that the sums drop out of Equation~\ref{sec:method:eqn:lrt}, leaving us with
\eqnstart\label{sec:method:eqn:bin-ts}
\begin{split}
D_b=&N_{1,T}\ln\left(1+\frac{M_{1,T}}{N_{1,T}}\right)+M_{1,T}\ln\left(1+\frac{N_{1,T}}{M_{1,T}}\right)\\
&-\sum_{i=1}^BN_{l_i,u_i}\ln\left(1+\frac{M_{l_i,u_i}}{N_{l_i,u_i}}\right)\\
&-\sum_{i=1}^BM_{l_i,u_i}\ln\left(1+\frac{N_{l_i,u_i}}{M_{l_i,u_i}}\right).
\end{split}
\eqnend
The total test statistic is simply the sum of the test statistics for the individual analysis bins:
\eqnstart\label{sec:method:eqn:full-ts}
D=\sum_bD_b.
\eqnend
Thus, $D$ is the log of the full likelihood ratio for any alternative hypothesis that divides the buffer into states of constant flux, compared to the null hypothesis of a single uniform flux state.

\subsection{Flare Monitor Trigger Condition}\label{sec:method:sub:trigger-condition}

The flare monitor searches for a single increased flux state in the buffer, which can be represented as an alternative hypothesis with $B=2$.
We allow the increased flux state to begin at any point $C$ in the buffer.
For convenience let us define $N\equiv N_{1,T}$, $M\equiv M_{1,T}$, $N_1\equiv N_{1,C-1}$, $M_1\equiv M_{1,C-1}$, $N_2\equiv N_{C,T}$, and $M_2\equiv M_{C,T}$.
That is, $N_1$ and $M_1$ are the sums of the on-source and off-source counts prior to the proposed increased flux state, and $N_2$ and $M_2$ are the respective sums during the proposed increased flux state.
Under these assumptions, Equation~\ref{sec:method:eqn:bin-ts} takes the form
\eqnstart\label{sec:method:eqn:bin-ts-trigger}
\begin{split}
D_b=&N\ln\left(1+\frac{M}{N}\right)+M\ln\left(1+\frac{N}{M}\right)\\
&-N_1\ln\left(1+\frac{M_1}{N_1}\right)-M_1\ln\left(1+\frac{N_1}{M_1}\right)\\
&-N_2\ln\left(1+\frac{M_2}{N_2}\right)-M_2\ln\left(1+\frac{N_2}{M_2}\right).
\end{split}
\eqnend

Since we are interested in issuing alerts only for flux increases, we require that our alternative hypothesis show an increase in the ratio $\rho$.
That is, if we define $\rho_1\equiv N_1/(\alpha M_1)$ and $\rho_2 \equiv N_2/(\alpha M_2)$, then we replace Equation~\ref{sec:method:eqn:full-ts} with
\eqnstart\label{sec:method:eqn:full-ts-trigger}
D=\sum_b\max\left[0,\sign(\rho_2-\rho_1)D_b\right],
\eqnend
where as before we have omitted the subscript $b$ on $\rho_1$ and $\rho_2$ for clarity.

The value of $D$ given by Equation~\ref{sec:method:eqn:full-ts-trigger} depends on $C$, the starting point of the hypothesized flare in the buffer.
The flare monitor determines the value of $D$ for each possible value of $C$, selecting the largest value, $D^\mathrm{max}$.
Due to the correlations involved in this search, we cannot easily derive the analytic distribution of $D^\mathrm{max}$ in the absence of any flares.
We elect instead to set a trigger threshold condition on the value of $D^\mathrm{max}$ and characterize the false alarm rate of the method as a function of this threshold.
Our trigger condition takes the form
\eqnstart\label{sec:method:eqn:trigger-condition}
D^\mathrm{max}>-\ln\gamma,
\eqnend
where $\gamma$ is a threshold parameter that fulfills $0<\gamma<1$.

\section{Tuning the False Alarm Rate}\label{sec:tuning}

The threshold parameter $\gamma$ allows us to control the false alarm rate of the flare detection method.
Under ideal circumstances, we would be able to calculate the false alarm rate analytically.
However, because the search is performed on a sliding time window, the large degree of correlation between searches prohibits such calculation.
We therefore estimate the false alarm rate by applying the search algorithm to simulated data.
To do this, we first determine the false alarm rate for sources located at the same declination as the Crab Nebula, which culminates at a zenith angle of $3^\circ$.
We then extend the simulation to other declinations in order to derive a declination-dependent correction factor for $\gamma$.
This correction brings the relationship between false alarm rate and threshold parameter into agreement at all declinations.
Following this correction, we identify data quality cuts necessary to resolve discrepancies between the simulated false alarm rate and the false alarm rate derived from data collected from regions of the sky that are at least $2^\circ$ away from any known VHE source or monitored target.
Finally, we discover that an additional correction to the threshold parameter is required for targets with strong steady VHE emission, on account of the small number of events in several of the analysis bins.

\subsection{Simulating the False Alarm Rate}\label{sec:tuning:sub:simulation}

We derive our simulated observations based on the observed off-source counts in the data.
The simulation thus mimics the rise and fall of the rates as the target rises and sets.
The only parameter that we require from the data is the observed off-source count $m_i$ in time bin $i$, for each analysis bin $b$.
After obtaining $m_i$ from the data, we sample a simulated background count $m_i'$ from a Poisson distribution with mean $m_i$ and a simulated source count $n_i'$ from a Poisson distribution with mean $\alpha m_i$.
This procedure enables us to simulate many thousands of years of data in order to understand precisely the relationship between the parameter $\gamma$ and the false alarm rate, while maintaining a realistic distribution of off-source counts as the target changes zenith angles while crossing the sky.

\figstart
\includegraphics[width=\figurewidth]{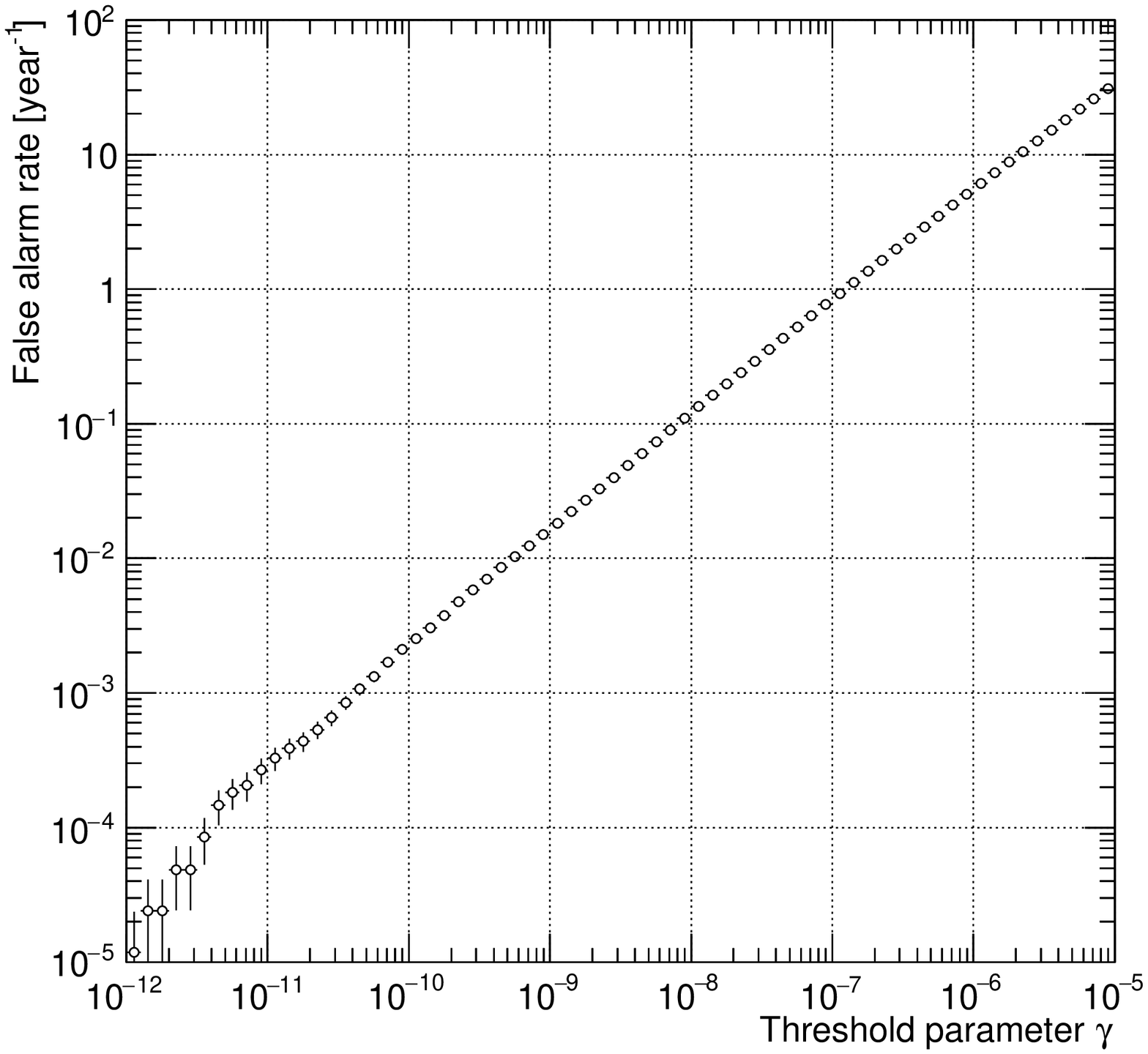}
\caption{
The cumulative false alarm rate as a function of the threshold parameter $\gamma$, derived from $2\times10^4$ years of simulated observations at the declination of the Crab Nebula.
}
\label{sec:tuning:fig:gamma-to-false-alarm-rate}
\figend

Since we sample $m_i'$ and $n_i'$ from the off-source distribution, the observed on-source distribution does not affect the simulation, so we can use data collected from regions with strong VHE sources as inputs to the simulation.
Figure~\ref{sec:tuning:fig:gamma-to-false-alarm-rate} shows the relationship between $\gamma$ and the false alarm rate derived from this method, using simulation sampled from Crab Nebula data.
We fit the points in Figure~\ref{sec:tuning:fig:gamma-to-false-alarm-rate} with a third-order spline in order to convert false alarm rate to threshold parameter.
The threshold parameters for false alarm rates of 1 event per target per year and 1 event per target per century are $\gamma=1.2\times10^{-7}$ and $\gamma=5.5\times10^{-10}$, respectively.

\subsection{False Alarm Rate Declination Dependence}\label{sec:tuning:sub:declination}

When a target is located at a large zenith angle, the on-source and off-source event counts are reduced compared to when the target is at a smaller zenith angle.
This is a result of the extensive air showers traversing a greater amount of atmosphere to reach the detector.
These smaller event counts make the data more likely to be fit well by a constant value, shifting the distribution of $D^\mathrm{max}$ to smaller values.
Since targets with declinations far from $+19^\circ$ spend more time at large zenith angles, we expect that the value of $\gamma$ required to meet the trigger condition of Equation~\ref{sec:method:eqn:trigger-condition} for a fixed false alarm rate will increase for these targets.
In this section, we derive a correction to the threshold parameter $\gamma$ that accounts for this effect.

\figstart
\includegraphics[width=\figurewidth]{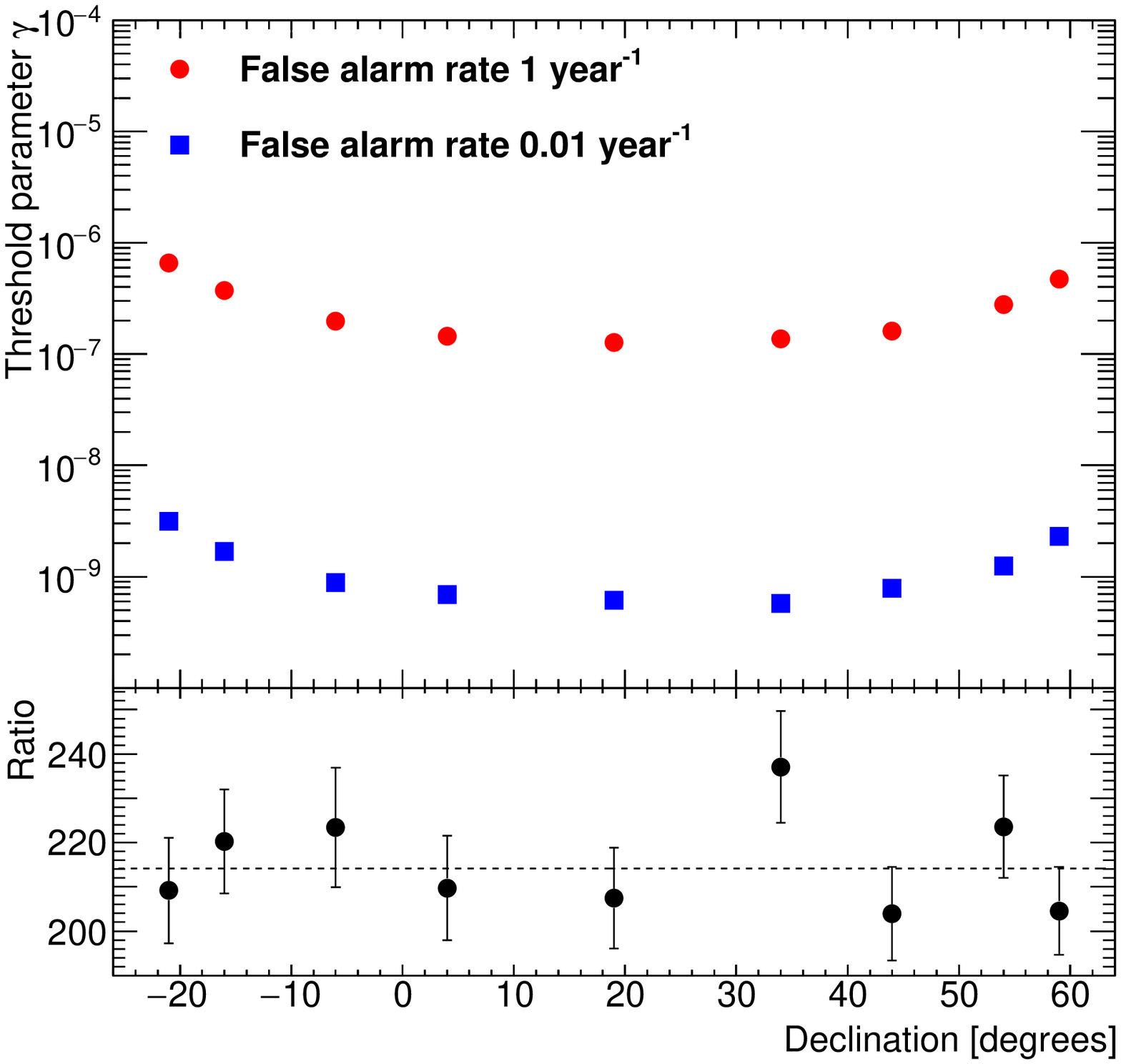}
\caption{
The threshold parameter as a function of declination.
The upper panel shows the threshold parameter required to achieve a false alarm rate of 1 event per target per year (red circles) and 1 event per target per century (blue squares).
All of the markers in the upper panel are larger than their errors.
In the lower panel, we depict the ratio of the threshold parameter for a false alarm rate of 1 event per target per year to that for a false alarm rate of 1 event per target per century, along with the best-fit constant value of 214.2 (dashed line).
}
\label{sec:tuning:fig:gamma-declination-correction}
\figend

Following the simulation procedure from Section~\ref{sec:tuning:sub:simulation}, we simulate results from nine different declinations, ranging from $-21^\circ$ to $+59^\circ$, in order to derive the declination dependence of the relationship between $\gamma$ and the false alarm rate.
The upper panel of Figure~\ref{sec:tuning:fig:gamma-declination-correction} shows the value of $\gamma$ required to obtain a fixed false alarm of 1 event per year and 1 event per century, as a function of declination.
The increase in $\gamma$ for declinations below $+4^\circ$ and above $+34^\circ$ demonstrates that our expectations are borne out by the simulation.

The lower panel of Figure~\ref{sec:tuning:fig:gamma-declination-correction} reveals that the ratio between the threshold parameters required to obtain these two false alarm rates does not change as a function of declination.
A fit of this ratio to a constant yields a value of 214.2, which describes the data well ($\chi^2$ of 7.2 for 8 degrees of freedom).
This independence of the ratio on declination allows us to apply a single declination-dependent correction to the threshold parameter that is constant for all false alarm rates.
We derive this correction from the results for a false alarm rate of 1 event per year.
After this correction is applied, the relationship between $\gamma$ and the false alarm rate follows the curve depicted in Figure~\ref{sec:tuning:fig:gamma-to-false-alarm-rate} for all targets, independent of the target declination.

\subsection{Data Quality Cuts}\label{sec:tuning:sub:quality}

Our implementation of the flare monitor algorithm is sensitive to changes that affect the detector response on time scales shorter than 4 hours, due to our use of the previous direct integration period to provide the off-source counts.
Although most detector maintenance leaves the response unchanged, occasionally the response shifts enough to affect the algorithm.
PMT rate spikes associated with lightning and similar random events can also affect the detector response.
Removing these unstable periods from the data processing is necessary for achieving the same false alarm rate in data that we obtain in simulation.
Since the flare monitor operates in real time, this requires a set of data quality cuts that can be applied automatically.

In order to assess the data quality, every 10 seconds we determine the overall event rate, the distribution of event zenith angles, and the two-dimensional angular distribution of events on the sky.
We compare the current distributions to the previous distributions, triggering a 2-hour pause in the accumulation of observations into the buffer if the rate changes by $\ge5$\%, or if the two-sample Kolmogorov-Smirnov probability for the angular distributions to be drawn from the same underlying distribution falls below $10^{-5}$.
The time interval of 2 hours is chosen in order to correspond to the direct integration time duration.
We find that applying these cuts retains 96\% of the data, showing that the detector is stable most of the time.

To verify the false alarm rates after applying the data quality cuts, we select a set of 150 regions at the declination of the Crab Nebula.
Each region is located at least $2^\circ$ away from any known VHE source or monitored target, in order to avoid possible contamination from actual flares.
We run the flare monitor algorithm over the data from these regions and compare the resulting false alarm rate to our simulation.

\figstart
\includegraphics[width=\figurewidth]{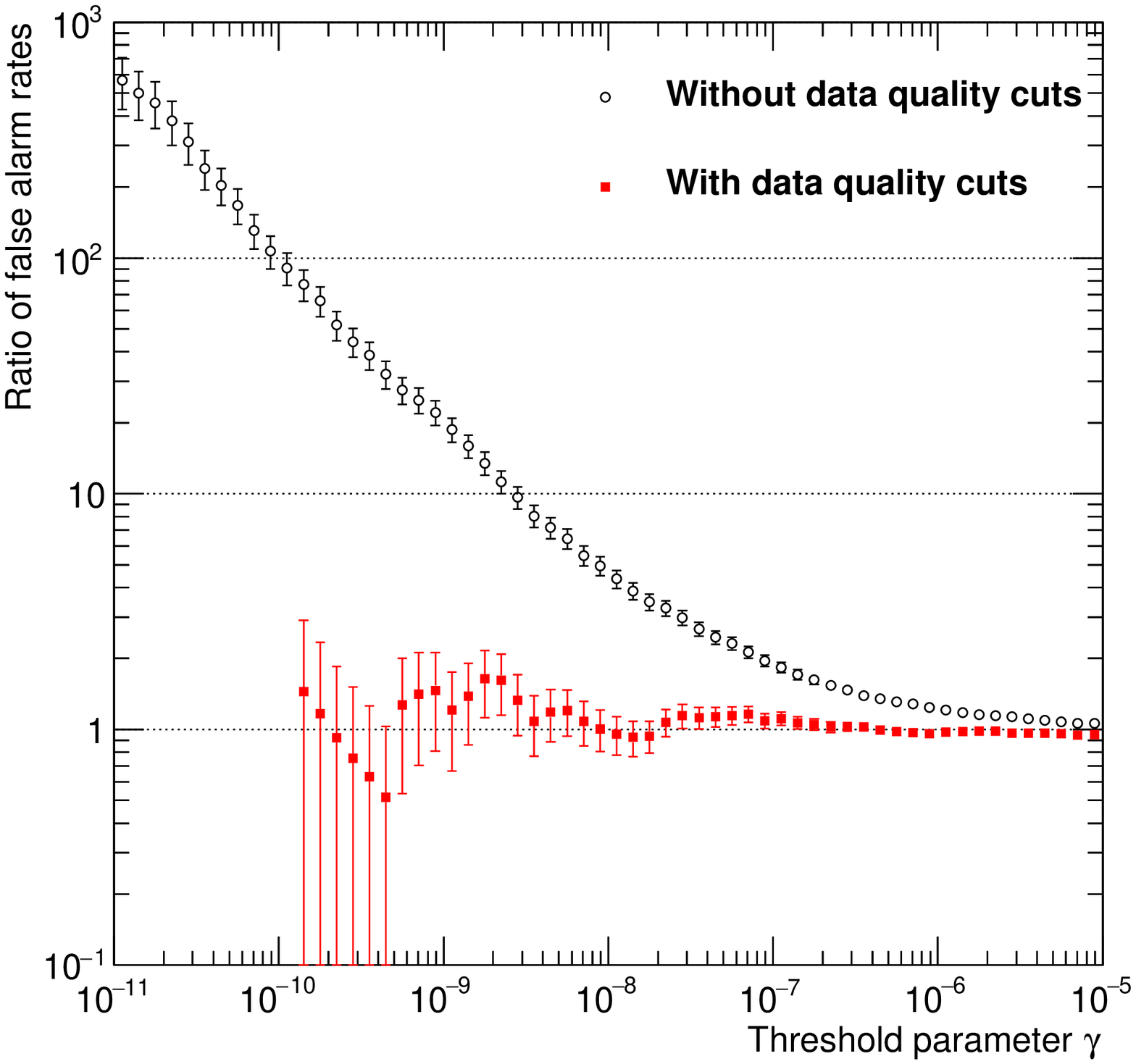}
\caption{
The ratio of the cumulative false alarm rate for the 150 regions discussed in the text to that expected from simulation, as a function of threshold parameter $\gamma$.
The open black circles represent the results without data quality monitoring, while the filled red squares include data suppression from 2 hours after any substantial change in the detector response.
}
\label{sec:tuning:fig:gamma-150-regions}
\figend

Figure~\ref{sec:tuning:fig:gamma-150-regions} shows the comparison between the false alarm rates in data and simulation.
Applying the data quality cuts removes events that would trigger with small threshold parameters.
These are events that appear to be highly significant, but actually arise due to detector instabilities.
As shown in Figure~\ref{sec:tuning:fig:gamma-150-regions}, all events that would trigger with a threshold parameter less than $10^{-10}$ are removed, and the observed false alarm rate matches the rate expected from simulation.
In contrast, in the absence of any data quality cuts, the false alarm rate derived from these 150 regions fails to match the simulated false alarm rate, becoming several orders of magnitude too large for small values of $\gamma$.
This mismatch reveals that our use of the previous direct integration period does indeed affect the data, necessitating our use of the data quality cuts.
Caution should be used in interpreting the error bars in Figure~\ref{sec:tuning:fig:gamma-150-regions} because the results presented are cumulative and therefore the points are correlated.

\subsection{Threshold Correction for Strong VHE Targets}\label{sec:tuning:sub:strong}

The higher analysis bins suffer from low count values.
Simulations reveal that this causes the distribution of the test statistic constructed from Equations~\ref{sec:method:eqn:bin-ts-trigger} and \ref{sec:method:eqn:full-ts-trigger} to have a weak dependence on the presence of steady VHE emission.
Targets with stronger steady emission tend to produce higher values of $D$ than targets with weaker or no steady emission in gamma rays, even in the absence of any flares.
Accounting for this effect is necessary to cause the false alarm rates for strong steady VHE emitters to match our expectations from Figure~\ref{sec:tuning:fig:gamma-to-false-alarm-rate}.

In order to quantify this effect, we perform simulations following the procedure outlined in Section~\ref{sec:tuning:sub:simulation}.
We account for the presence of steady emission by including an additional term in the mean of the Poisson distribution from which the on-source counts are drawn.
Instead of using a mean of $\alpha m_i$ for this distribution, which models a source region that has only background counts, we use a mean of $\alpha m_i+r_\textnormal{target}\alpha m_i$, where $r_\mathrm{target}$ is the long-term relative excess observed at the target location:
\eqnstart\label{sec:tuning:eqn:relative-excess}
r_\mathrm{target}=\frac{n_\mathrm{target}-\alpha m_\mathrm{target}}{\alpha m_\mathrm{target}}.
\eqnend
Here, $n_\mathrm{target}$ and $m_\mathrm{target}$ are the total on-source and off-source counts collected from the vicinity of the strong target over the entire data set, yielding a highly accurate estimate of $r_\mathrm{target}$.
Thus, the additional term of the relative excess $r_\mathrm{target}$ multiplied by $\alpha m_i$ represents the excess expected from the source.
The values of $n_\mathrm{target}$, $m_\mathrm{target}$, and $r_\mathrm{target}$ depend on the analysis bin.

\figstart
\includegraphics[width=\figurewidth]{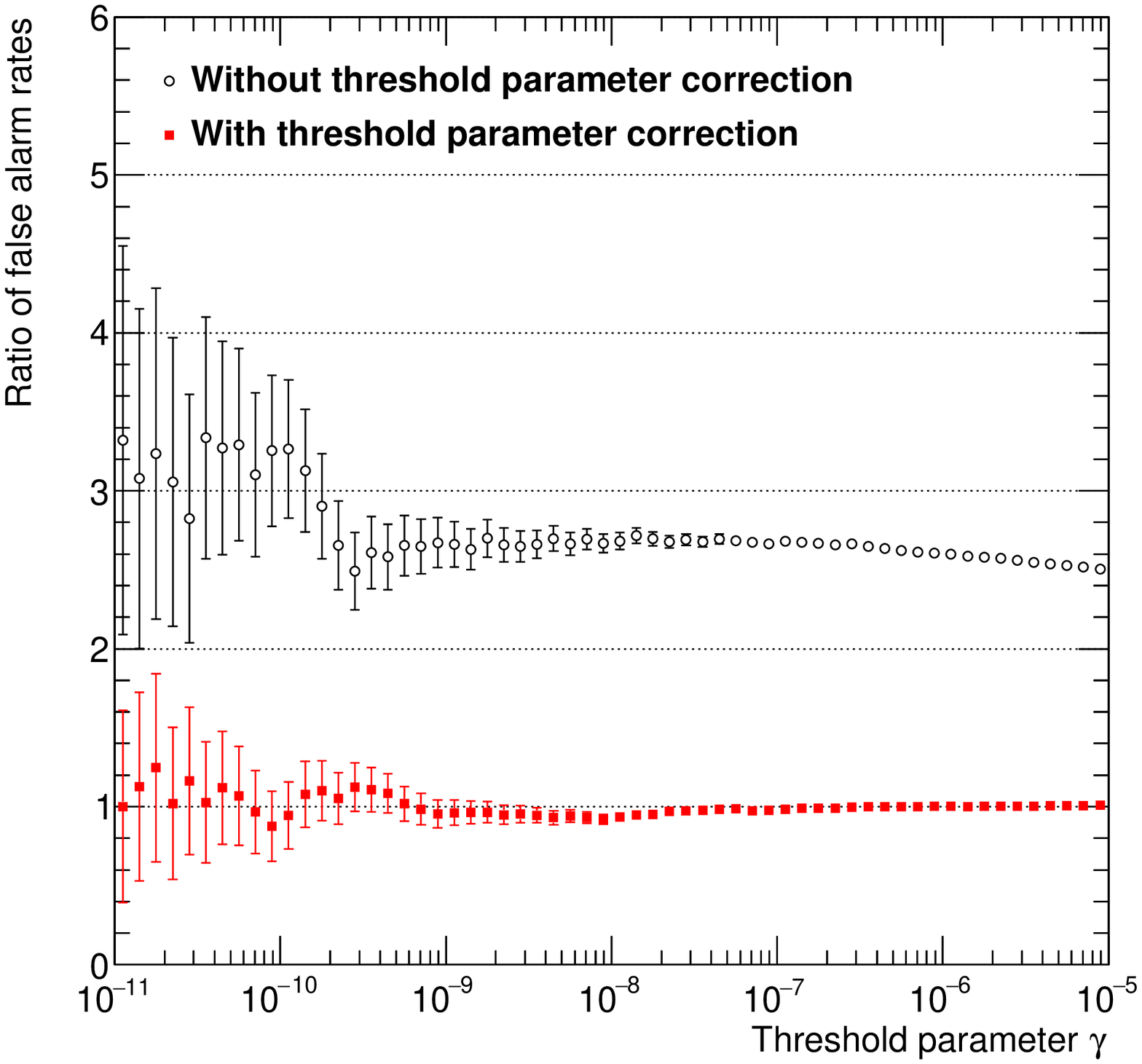}
\caption{
The ratio of the cumulative false alarm rate from the simulated Crab to that expected from a source with no steady VHE emission.
The open black circles correspond to points without any additional correction to the threshold parameter.
A correction factor of $k_\mathrm{Crab}=1.2$ (see Equation~\ref{sec:tuning:eqn:trigger-correction}) brings the false alarm rate into agreement with the expectations for a source with no steady VHE emission (filled red squares).
}
\label{sec:tuning:fig:simcor}
\figend

Figure~\ref{sec:tuning:fig:simcor} depicts the cumulative ratio $R_\mathrm{Crab,sim}/R_\mathrm{0,sim}$ as a function of $\gamma$, where $R_\mathrm{Crab,sim}$ is the simulated false alarm rate for a source with steady VHE emission equal to that observed from the Crab Nebula, and $R_\mathrm{0,sim}$ is the simulated false alarm rate in the absence of any VHE source as shown in Figure~\ref{sec:tuning:fig:gamma-to-false-alarm-rate}.
Thus, $R_\mathrm{Crab,sim}(\gamma')$ and $R_\mathrm{0,sim}(\gamma')$ are the rates at which false alarms occur when the threshold parameter is set to $\gamma=\gamma'$.
The open black circles in Figure~\ref{sec:tuning:fig:simcor} show that the false alarm rate for a Crab-like source is $\sim$2.5 times higher than what we would expect in the absence of any source.
We combat this effect by introducing a target-dependent correction factor $k_\mathrm{target}$ to the trigger condition expressed by Equation~\ref{sec:method:eqn:trigger-condition}, which now takes the form
\eqnstart\label{sec:tuning:eqn:trigger-correction}
D^\mathrm{max}>-\ln\gamma+k_\mathrm{target}.
\eqnend
For $k_\mathrm{target}>0$, this requires a larger value of $D^\mathrm{max}$ in order for the algorithm to trigger than would have been required in the absence of any steady VHE emission.
Using a correction factor of $k_\mathrm{Crab}=1.2$ for a source with Crab-like VHE emission results in the filled red squares shown in Figure~\ref{sec:tuning:fig:simcor}.
These points demonstrate that a single correction factor suffices to bring the false alarm rate into agreement with simulation for all values of the threshold parameter $\gamma$.

The value of $k_\mathrm{target}$ is different for every VHE source.
As outlined in Section~\ref{sec:targets}, the only strong sources in our current target list are the blazars Mrk 421 and Mrk 501.
These sources are generally weaker than the Crab, and the above simulation procedure produces a value of $k_\mathrm{Mrk}=0.2$ for both Mrk 421 and Mrk 501.
For all other targets, we use a value of $k_\mathrm{target}=0$.

\section{Sensitivity of the Flare Monitor}\label{sec:sensitivity}

\editadded{
We next investigate the sensitivity of the flare monitor to flares with various properties.
The flux, duration, structure, and source declination of the flare can all affect both the probability of detection and the speed with which the flare is detected.
}

\subsection{\editadded{Flare Simulation Procedure}}\label{sec:sensitivity:sub:simulation}

We characterize the sensitivity of the flare monitor by injecting fiducial flares represented by a constant increase in flux for a duration of 1 hour with the same spectral shape as that of the Crab Nebula as measured by HAWC.
We adapt the procedure outlined in Section~\ref{sec:tuning:sub:strong}, sampling the simulated off-source count $m_i'$ from a Poisson distribution with mean $m_i$ and the simulated on-source count $n_i'$ before and after the flare from a Poisson distribution with mean $\alpha m_i$.
During the flare, we adjust the mean of the on-source count distribution to $(1+fr_\mathrm{Crab})\alpha m_i$, where $r_\mathrm{Crab}$ is the observed relative excess of events collected from the vicinity of the Crab Nebula as defined by Equation~\ref{sec:tuning:eqn:relative-excess} and $f$ is the flare flux in Crab units.
Although the relative excess $r_\mathrm{Crab}$ depends on the analysis bin, in our fiducial case the flare flux $f$ does not.
The spectrum of the simulated flares therefore differs from that of the Crab Nebula only by the overall constant $f$.
We assume no underlying steady state emission for the source, so that the analysis presented here is applicable to targets with little or no VHE emission.
The results are equally applicable to strong VHE emitters, provided that correction factors are derived according to the procedures outlined in Sections~\ref{sec:tuning:sub:declination} and \ref{sec:tuning:sub:strong}.

Similarly to the tuning of the analysis bins referred to in Section~\ref{sec:hawc-operation}, we use the measured spectrum of the Crab Nebula as the input flare spectrum.
Blazar flare spectra are likely to be harder than the Crab, but they are also affected by EBL absorption.
A thorough assessment of the sensitivity of the flare monitor to a range of different intrinsic spectra and source redshifts is beyond the scope of this work.
Further studies that address the effects of more realistic flare spectra are currently in progress.

We inject the simulated flares uniformly randomly in time, independent of whether the target spends any time within the field of view throughout the duration of the flare.
We consider a flare to be detectable in principle if any part of the flare occurs while the source is at zenith angles smaller than the $45^\circ$ zenith angle cut on the events.
As a result, we do not expect the probability of detection to converge to 1, since some flares will pass this cut only for a small fraction of their duration.

The flares are separated by an amount of time sufficient to fill the entire 10-hour buffer in order to guarantee that only one flare exists in the buffer at any given time.
This procedure gives us the sensitivity to isolated flares.
We run the algorithm detailed in Section~\ref{sec:method} over the simulated data set.
Each time the buffer advances forward in time, we evaluate Equations~\ref{sec:method:eqn:bin-ts-trigger} and \ref{sec:method:eqn:full-ts-trigger} and apply the trigger condition expressed by Equation~\ref{sec:method:eqn:trigger-condition} to determine whether the flare is detected.
We generate a sufficient number of simulated data, 300 observations of 2 minutes' duration each, to fill the buffer before the flare starts, and we continue to generate data until the flare has completely exited the buffer.
This procedure insures that the probability to detect the flare is unaffected by the presence of other flares.

\subsection{\editadded{Sensitivity to Flare Strength}}\label{sec:sensitivity:sub:simulation}

\figstart
\includegraphics[width=\figurewidth]{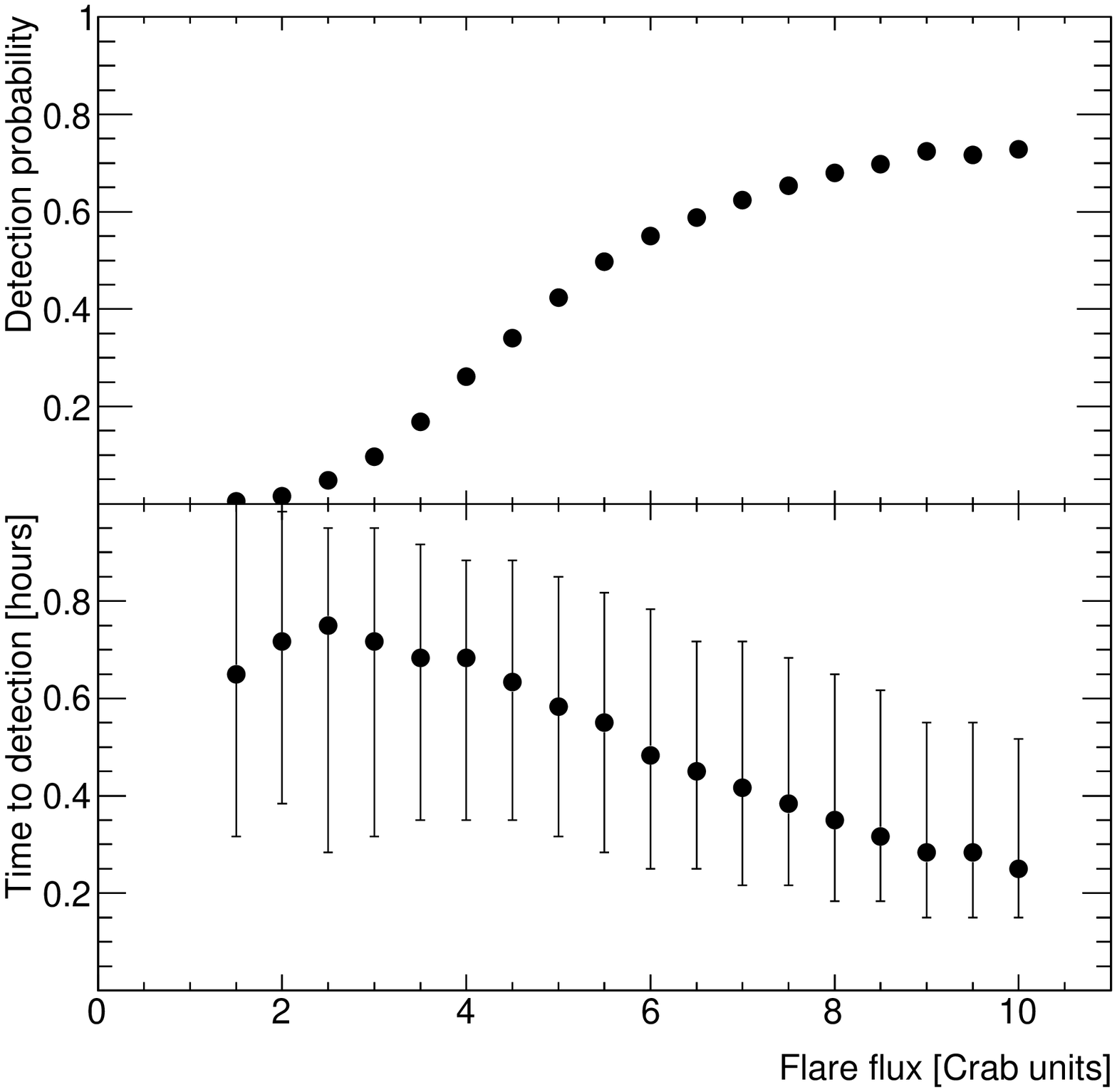}
\caption{
Sensitivity of the flare monitor to fiducial flares injected into the simulated data, as a function of flare flux in Crab units.
The upper panel shows the fraction of detectable flares identified by the trigger algorithm.
In the lower panel, the points represent the median time to detection for flares that are successfully identified.
The error bars depict the central $68\%$ of the distribution of times to detection for the flares.
}
\label{sec:sensitivity:fig:base-sensitivity}
\figend

Figure~\ref{sec:sensitivity:fig:base-sensitivity} presents the sensitivity of the flare monitor to fiducial flares of varying fluxes.
The upper panel of the figure shows the detection probability as a function of flare flux.
The data points are derived from observations at the declination of the Crab Nebula, and the false alarm rate is set to 10 events per year from the entire target collection defined in Section~\ref{sec:targets}.
Since the flares are injected at uniformly random times, the detection probability for strong flares converges to a value somewhat less than 1 because flares that occur at the edges of the transits are counted as detectable, even though their lower event counts make their detection less likely.
The probability of detecting a given flare reaches $50\%$ at approximately 5.5 Crab units.

The lower panel of Figure~\ref{sec:sensitivity:fig:base-sensitivity} shows the time to detection---the amount of time between the start of the flare and the time at which the trigger is issued---as a function of flare flux.
The points in the lower panel of the figure show the median time to detection, while the error bars represent the central $68\%$ of the distribution.
This distribution is wide partly because flares occurring near target culmination are detected more quickly than flares occurring when the target is at larger zenith angles.
Of the flares that are detected, the majority are identified in less than their duration of 1 hour, demonstrating that the method can trigger sufficiently rapidly to enable follow-up observations.

\subsection{\editadded{Sensitivity to Flare Duration}}\label{sec:sensitivity:sub:simulation}

\figstart
\includegraphics[width=\figurewidth]{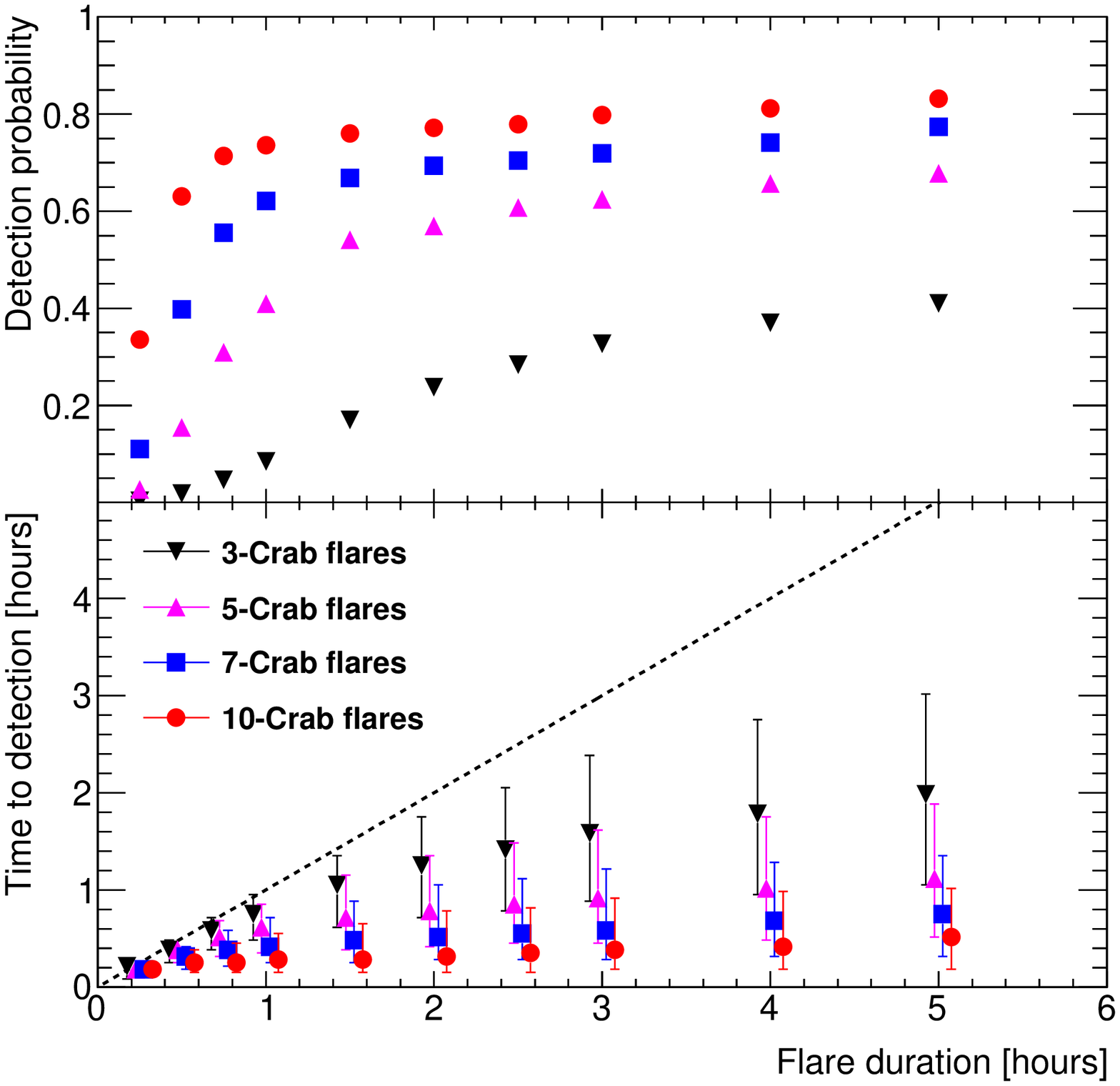}
\caption{
Sensitivity of the flare monitor to flares of varying duration.
The panels are the same as in Figure~\ref{sec:sensitivity:fig:base-sensitivity}, except that they now show the flare detection probability (top) and time to detection (bottom) as a function of flare duration instead of flare flux.
Different points show the results for flares with fluxes equal to 3, 5, 7, and 10 times the flux from the Crab Nebula.
The dashed line in the lower panel shows the points at which the time to detection is equal to the flare duration.
Points below this line are detected before the flare ends.
The points in the lower panel are slightly offset from their exact durations for clarity.
}
\label{sec:sensitivity:fig:duration-sensitivity}
\figend

Figure~\ref{sec:sensitivity:fig:duration-sensitivity} expands the results of Figure~\ref{sec:sensitivity:fig:base-sensitivity} to flares of durations other than 1 hour, for flares with fluxes equal to 3, 5, 7, and 10 times the Crab Nebula flux.
As before, the upper panel shows the detection probability as a function of flare duration, while the lower panel shows the time to detection.
An increase in flare duration improves the detection probability for a fixed flare flux.
For instance, whereas Figure~\ref{sec:sensitivity:fig:base-sensitivity} shows that the probability of detecting a 3-Crab flare that lasts for 1 hour is $\sim$10\%, Figure~\ref{sec:sensitivity:fig:duration-sensitivity} reveals that this probability rises to $\sim$40\% if the duration increases to 5 hours.
Most importantly, the lower panel of Figure~\ref{sec:sensitivity:fig:duration-sensitivity} shows that the time to detection rises more slowly than the flare duration.
Thus, long-duration flares provide a larger window of time for other instruments to conduct follow-up observations before the flare terminates.

\subsection{\editadded{Sensitivity to Flare Structure}}\label{sec:sensitivity:sub:simulation}

\figstart
\includegraphics[width=\figurewidth]{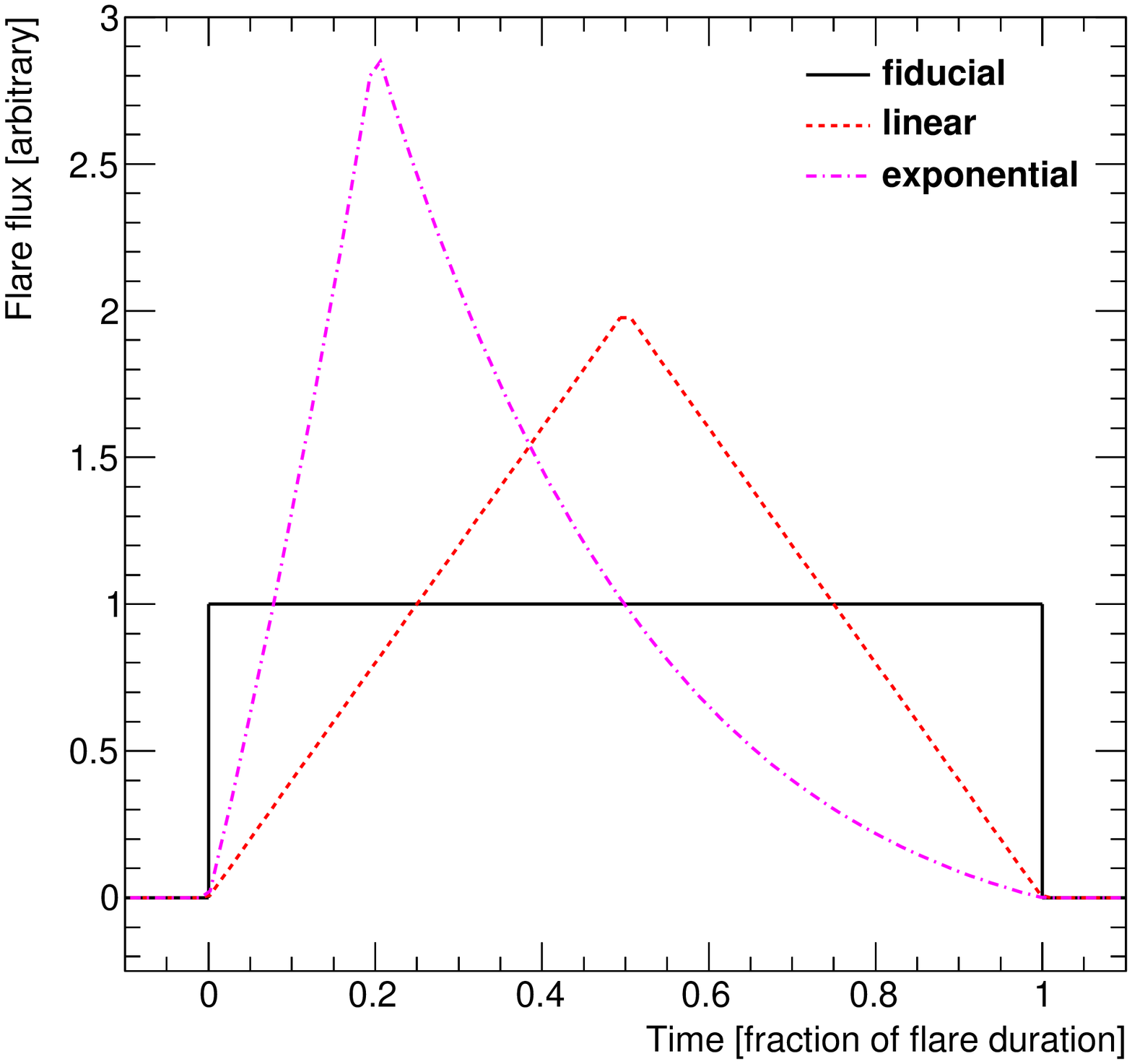}
\caption{
\editadded{
Flux versus time for three different flare shapes discussed in the text: the fiducial square-wave flare (solid black line), the linear flare with symmetric rise and fall (dashed red line), and the exponential fast rising and slowly falling flare (dot-dashed magenta line).
Time is measured as a fraction of total flare duration.
The flares are normalized to have the same integral.
}
}
\label{sec:sensitivity:fig:flare-shapes}
\figend

While our fiducial flare comprises no internal structure aside from a constant elevated flux, many detected VHE flares exhibit a large degree of structure in their light curves.
We have tested \editremoved{several} \editadded{three} different light curve shapes\editadded{, shown in Figure~\ref{sec:sensitivity:fig:flare-shapes}}, to ascertain how the flare monitor sensitivity may change in the presence of such structure.
\editadded{In addition to the fiducial square wave flare, we have also tested a flare that rises linearly for half of the flare duration and falls linearly for the other half (hereafter referred to as the linear case), as well as a flare that rises exponentially over the first fifth of the flare duration and then falls exponentially over the remainder (hereafter referred to as the exponential case).}
\editremoved{Specifically, we have tested linear, quadratic, and Gaussian dependences on time.
In general, for a fixed fluence---total particle flux integrated over the time of the flare---the presence of structure in the flare increases the detection probability, while the nature of the structure determines the speed with which the flares are detected.
Flares that increase slowly require a larger time to detection, whereas flares that rise quickly and fade away slowly are detected more rapidly.
Since the presence of structure increases the detection probability, the sensitivity plots presented in this section may be regarded as lower limits on the detection probability, subject to the presence of internal structure in the light curve of the flare.
}

\figstart
\includegraphics[width=\figurewidth]{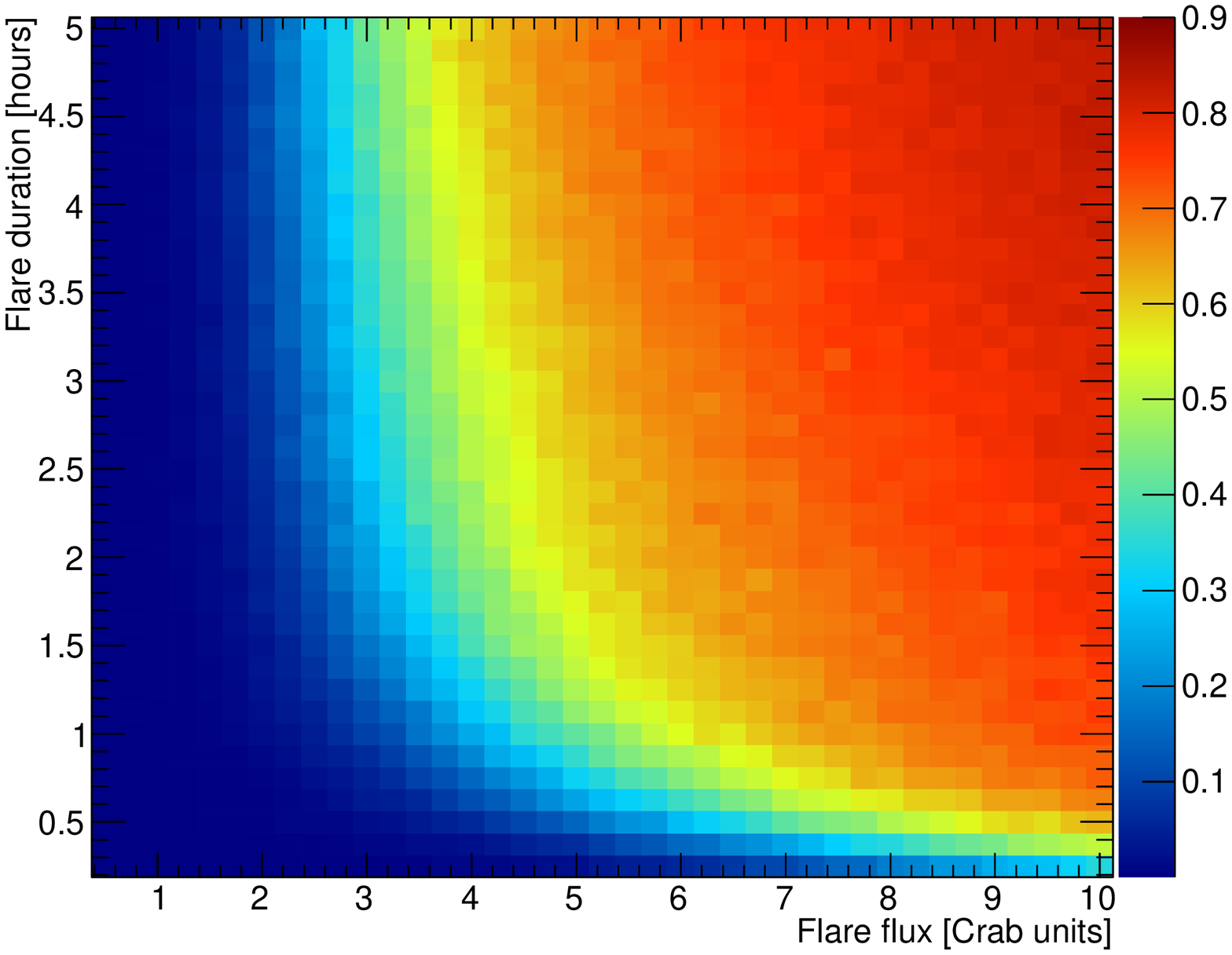}
\caption{
\editadded{
Fraction of detectable fiducial flares identified by the trigger algorithm as a function of average flare flux and flare duration.
}
}
\label{sec:sensitivity:fig:flare-fiducial2d}
\figend

\editadded{
Figure~\ref{sec:sensitivity:fig:flare-fiducial2d} shows the detection probability for the fiducial flare as a function of both flare flux and flare duration.
In the figure, the flare flux is the average flux over the entire flare duration, which in the fiducial case is equivalent to the constant flux of the flare.
As expected, increasing either the flux or the duration improves the detection probability.
}

\figstart
\includegraphics[width=\figurewidth]{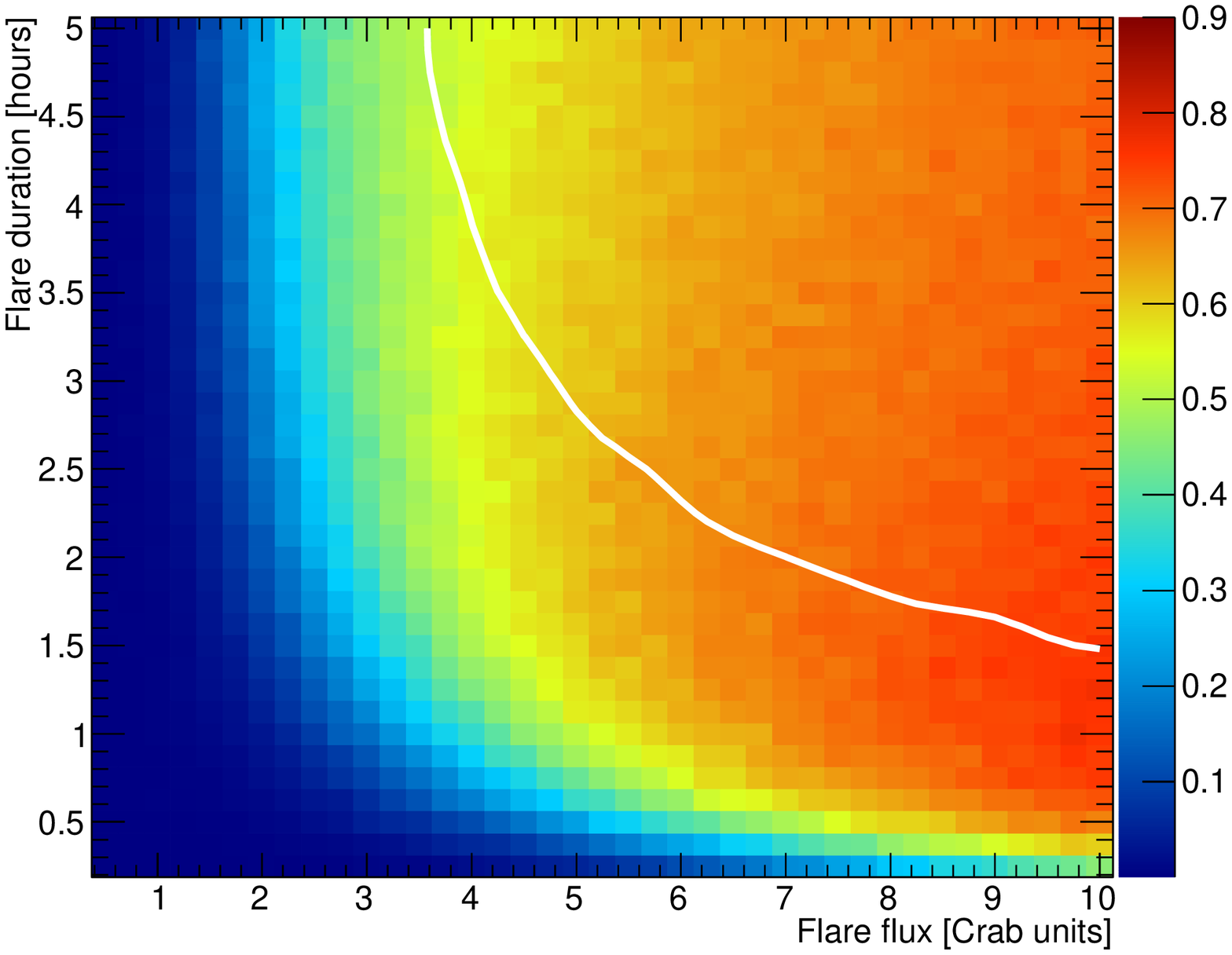}
\caption{
\editadded{
Fraction of detectable linear flares (see Figure~\ref{sec:sensitivity:fig:flare-shapes} for the flare shape) identified by the trigger algorithm as a function of integrated flare flux and flare duration.
Flares lying above the white contour are more likely to be detected in the fiducial flare case, while flares below the contour are more likely to be detected when the flare structure is linear.
}
}
\label{sec:sensitivity:fig:flare-linear2d}
\figend

\editadded{
The detection probability for the linear flare structure appears in Figure~\ref{sec:sensitivity:fig:flare-linear2d}.
As in the fiducial case, the flare flux is the average over the duration of the flare, which in the linear case is half of the peak flux.
Points in Figure~\ref{sec:sensitivity:fig:flare-linear2d} therefore represent points with the same fluence---total flux integrated over the duration of the flare---as the corresponding points in Figure~\ref{sec:sensitivity:fig:flare-fiducial2d}.
}

\editadded{
In the linear case, the probability tends to peak for flares of around 1.5 hours' duration and gradually decreases for flares that last longer than this.
This trend is most apparent for the strongest flares, and can be understood as an effect of the source transit.
Since we inject flares uniformly randomly in time, a fraction of flares will occur near the edges of the transit, as the source enters or leaves the field of view.
We therefore sample only the beginning or end of these truncated flares.
For such flares, a smaller fraction of the total fluence from linear flares appears in the field of view than in the fiducial case.
When the flare duration is small compared to the transit duration of $\sim$6 hours, the fraction of truncated flares is negligible.
However, as the flare duration increases, a larger fraction of flares are subjected to this effect, which manifests itself as a reduction in the probability of detection for a fixed total fluence.
The white contour in Figure~\ref{sec:sensitivity:fig:flare-linear2d} shows the point at which the probability of flare detection in the linear case drops below that of the fiducial case as a result of this effect.
Overall, the linear structure does not reduce the detection probability by more than $10\%$, and for weak or short flares the structure increases the detection probability by as much as a factor of 2.
This suggests that light-curve structure, which is expected to be present in most flares, will enhance the detection probability in many cases.
}

\editadded{
We have also tested the sensitivity of our method to flares with the exponential profile from Figure~\ref{sec:sensitivity:fig:flare-shapes}.
The results are very similar to those for the linear case and we consequently do not discuss them here.
}

\subsection{\editadded{Sensitivity to Flare Source Declination}}\label{sec:sensitivity:sub:simulation}

\figstart
\includegraphics[width=\figurewidth]{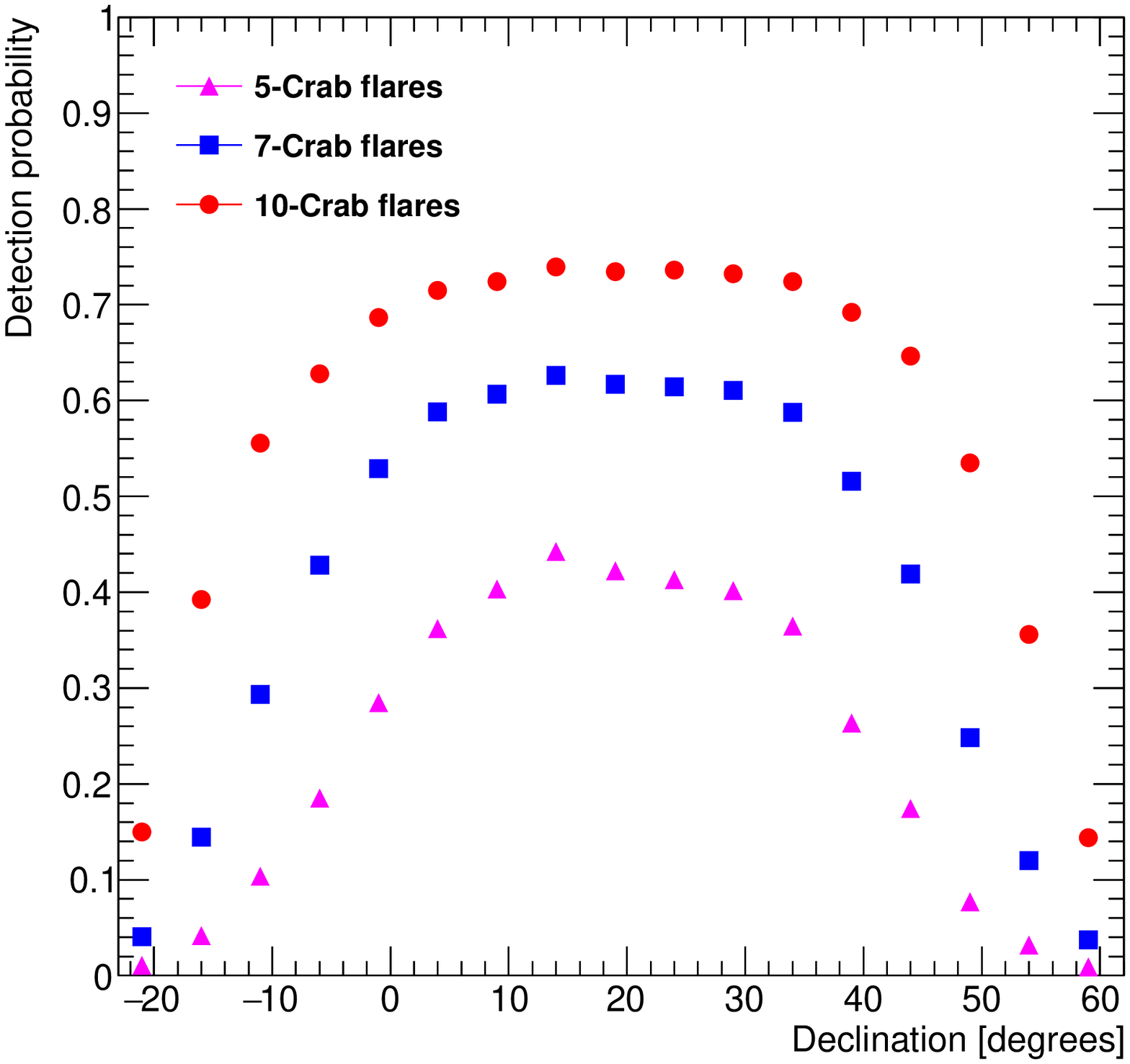}
\caption{
Declination dependence of the flare monitor sensitivity for fiducial flares with fluxes equal to 5, 7, and 10 times the flux from the Crab Nebula.
}
\label{sec:sensitivity:fig:declination-sensitivity}
\figend

The sensitivity depends strongly on the declination of the target, primarily due to the reduced number of events at large zenith angles, which renders discrimination of elevated flux states from the background rate increasingly difficult.
Figure~\ref{sec:sensitivity:fig:declination-sensitivity} depicts the sensitivity to flares from targets at a range of declinations.
The figure is constructed for the same false alarm rate that was used to construct Figure~\ref{sec:sensitivity:fig:base-sensitivity}.
This rate accounts for the declination-dependent threshold parameter correction shown in Figure~\ref{sec:tuning:fig:gamma-declination-correction}.
The asymmetry apparent in Figure~\ref{sec:sensitivity:fig:declination-sensitivity} arises due to HAWC being located in the northern hemisphere.
At declinations of $-16^\circ$ and $+54^\circ$, located $35^\circ$ away from the latitude of the HAWC observatory, the probability of detecting a flare of 10 Crab units is small but non-negligible.
We monitor targets at a maximum of $40^\circ$ away from the HAWC latitude in order to provide sensitivity to these extreme events.
Since we cut events with zenith angles above $45^\circ$, these sources are monitored for much less time per day than sources located nearer to the HAWC latitude.

\section{Target Selection}\label{sec:targets}

In the ideal case, we would be able to tile the sky with targets for the search algorithm and perform all-sky monitoring for VHE transients.
However, achieving an acceptably low false alarm rate under this approach requires us to lower the value of the threshold parameter $\gamma$, in turn lowering the sensitivity of the search.
For this reason, we concentrate on identifying a specific list of the most promising candidates, reserving an all-sky search for future work.

We select targets from two source catalogs: the online catalog of known VHE sources, TeVCat\footnote{See \url{http://tevcat.uchicago.edu}.} \citep{Wakely:2008vy}; and the second catalog of hard-spectrum Fermi-LAT sources, or 2FHL \citep{Ackermann:2016gt}.
The former catalog comprises sources that have been observed to produce VHE emission, while the latter represents sources strongly detected by the Fermi-LAT above 50 GeV that consequently have a large probability to produce VHE-band emission.
Although some Galactic sources, such as binaries, are known to flare in the VHE band \citep{Aharonian:2005ey,Albert:2006eu}, for this work we restrict our attention to extragalactic objects because we regard these as being far more likely to produce strong VHE transients.

Both catalogs also contain several unidentified sources with no known or obvious counterpart at other wavelengths.
Since the TeVCat unidentified sources cluster along the Galactic Plane, we exclude them from our target list as being likely Galactic in origin.
In contrast, the contribution of extragalactic sources to the 2FHL unidentified source population is likely $\sim$80\% \citep{Ackermann:2016gt}, so we include these objects in our analysis.
We also exclude targets culminating at a zenith angle of greater than $40^\circ$ in order to limit events to the region where HAWC is sensitive and the gamma-hadron response is stable.
For the latitude of HAWC, we therefore accept target locations between $-21^\circ$ and $+59^\circ$ in declination.

\begin{deluxetable*}{rcc}
\tablecaption{Definition of Target Categories\label{sec:targets:tab:target-categories}}
\tablehead{
\colhead{Category} & \colhead{Targets} & \colhead{\editadded{Threshold Parameter $\gamma$}}
}
\startdata
Mrk 421 \& Mrk 501 & 2 & \editadded{$1.6\times10^{-7}$} \\
Other TeVCat Extragalactic & 44 & \editadded{$4.2\times10^{-9}$} \\
Nearby 2FHL Extragalactic ($z<0.3$) & 22 & \editadded{$9.3\times10^{-9}$} \\
Other 2FHL Extragalactic \& Unknown & 119 & \editadded{$1.3\times10^{-9}$} \\
\tableline
Total & 187 \\
\enddata
\tablecomments{\editadded{The threshold parameter $\gamma$, defined in Equation~\ref{sec:method:eqn:trigger-condition}, controls the false alarm rate of the flare detection algorithm.}}
\end{deluxetable*}

After these cuts, our final selection comprises a total of 187 targets: 46 TeVCat sources and 141 additional 2FHL sources that are not in TeVCat.
To enhance the algorithm's sensitivity to targets that are more likely to produce transients, we divide the targets into four categories, defined in Table~\ref{sec:targets:tab:target-categories}.
We split the TeVCat sources into two categories: one containing the very strong VHE emitters Mrk 421 and Mrk 501, and another containing everything else.
We also split the 2FHL sources into two categories: one comprising sources with measured redshifts less than 0.3, because these sources are less likely to be affected by gamma-ray attenuation by the EBL; and a separate category covering more distant sources, sources without a measured redshift, and the unidentified sources.
We then choose the threshold parameter $\gamma$ independently for each category, such that each category contributes 1/4 of the total expected false alarm rate.
Thus, we choose a less restrictive value of $\gamma$, for instance, for the nearby 2FHL targets, reflecting our opinion that these targets are more likely to produce flares than the rest of the 2FHL sources, and their smaller number allows us to increase our sensitivity to them.
\editadded{The specific value of the threshold parameter chosen for each category appears in the third column of Table~\ref{sec:targets:tab:target-categories}.}
Of course, we could have chosen to divide the false alarm rate contributions differently among the target categories, but since this is degenerate with the definitions of the target categories themselves, our arbitrary choice of an even division is unimportant.

\section{Verification of the Method with Data}\label{sec:verify}

As a verification of the search method, we process archival HAWC data collected between 2014 November 26 and 2016 June 1 for Mrk 421 and Mrk 501, which are known to produce frequent strong flares.
In this data set, we identify six events surpassing the threshold that would have been set to achieve a false alarm rate of 10 events per year for the entire source selection described in Section~\ref{sec:targets}.
This section first takes a detailed look at the flare monitor's view of the strongest of these events, which is associated with Mrk 501.
We then summarize the rest of the triggers, reserving a thorough investigation of these events and the remainder of the targets for future work.

\subsection{Markarian 501 Flare on 2015 August 18}\label{sec:verify:sub:flare57252}

\figstart
\includegraphics[width=\figurewidth]{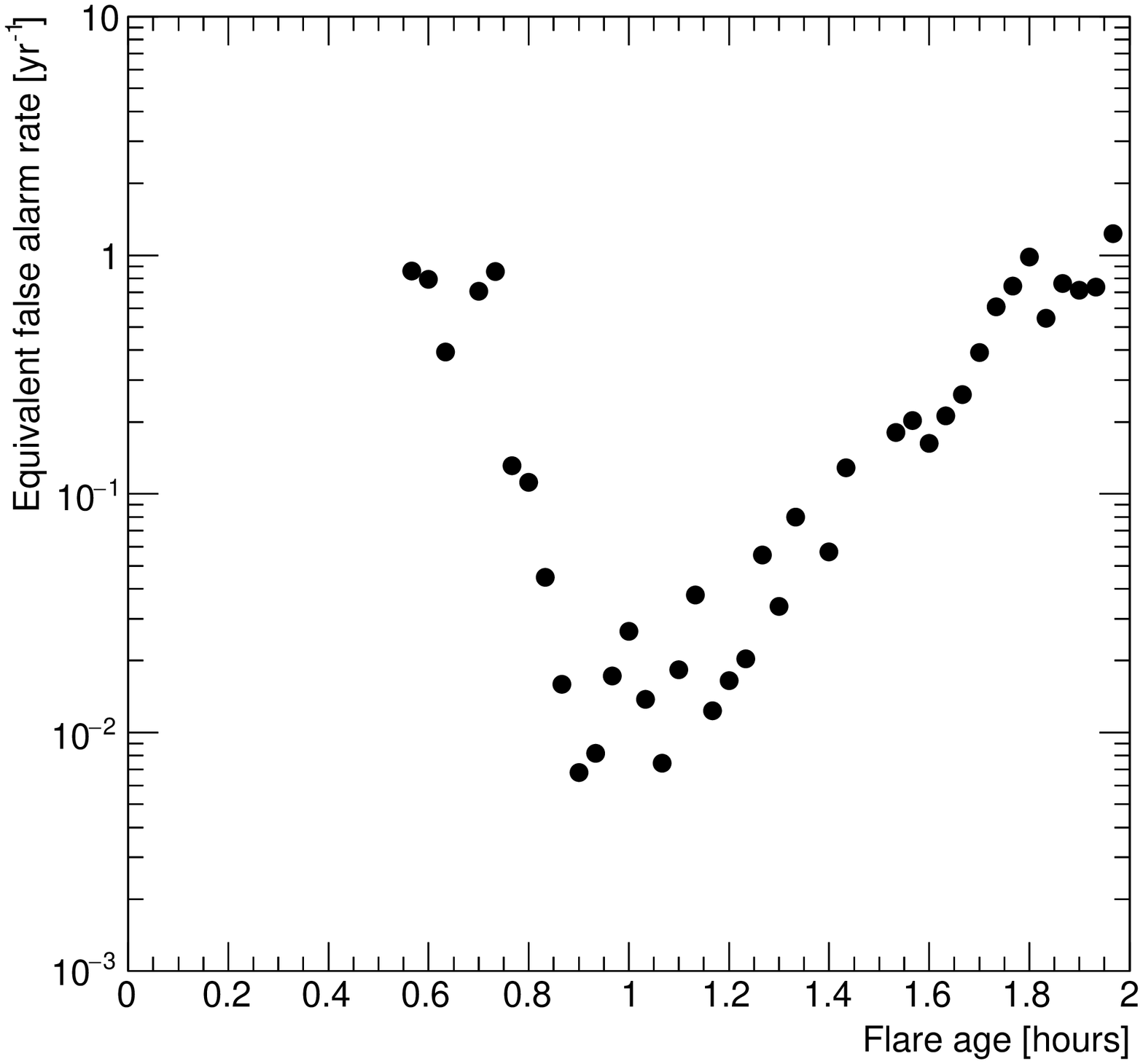}
\caption{
Equivalent false alarm rate versus flare age for the Mrk 501 flare detected on MJD 57252.
}
\label{sec:verify:fig:mrk501-flare57252-far}
\figend

The strongest of the six events identified by the flare monitor in the Mrk 421 \& Mrk 501 target class is associated with Mrk 501.
It occurred on 2015 August 18, at modified Julian day (MJD) 57252.062.
This event crossed the trigger threshold for the Markarian target class after 34 minutes.
Figure~\ref{sec:verify:fig:mrk501-flare57252-far} shows the equivalent false alarm rate of this event as a function of the age of the flare, which is the amount of time after the trigger time MJD 57252.062.
The minimum equivalent false alarm rate occurs 54 minutes after the identified start time of the event, achieving a value of $6.8\times10^{-3}$ events per year.
In other words, in the absence of any flares, we would expect to see an event at least this extreme only once every $\sim$150 years from a single source, or once every $\sim$75 years from the Mrk 421 \& Mrk 501 target class.

\wfigstart
\includegraphics[width=\wfigurewidth]{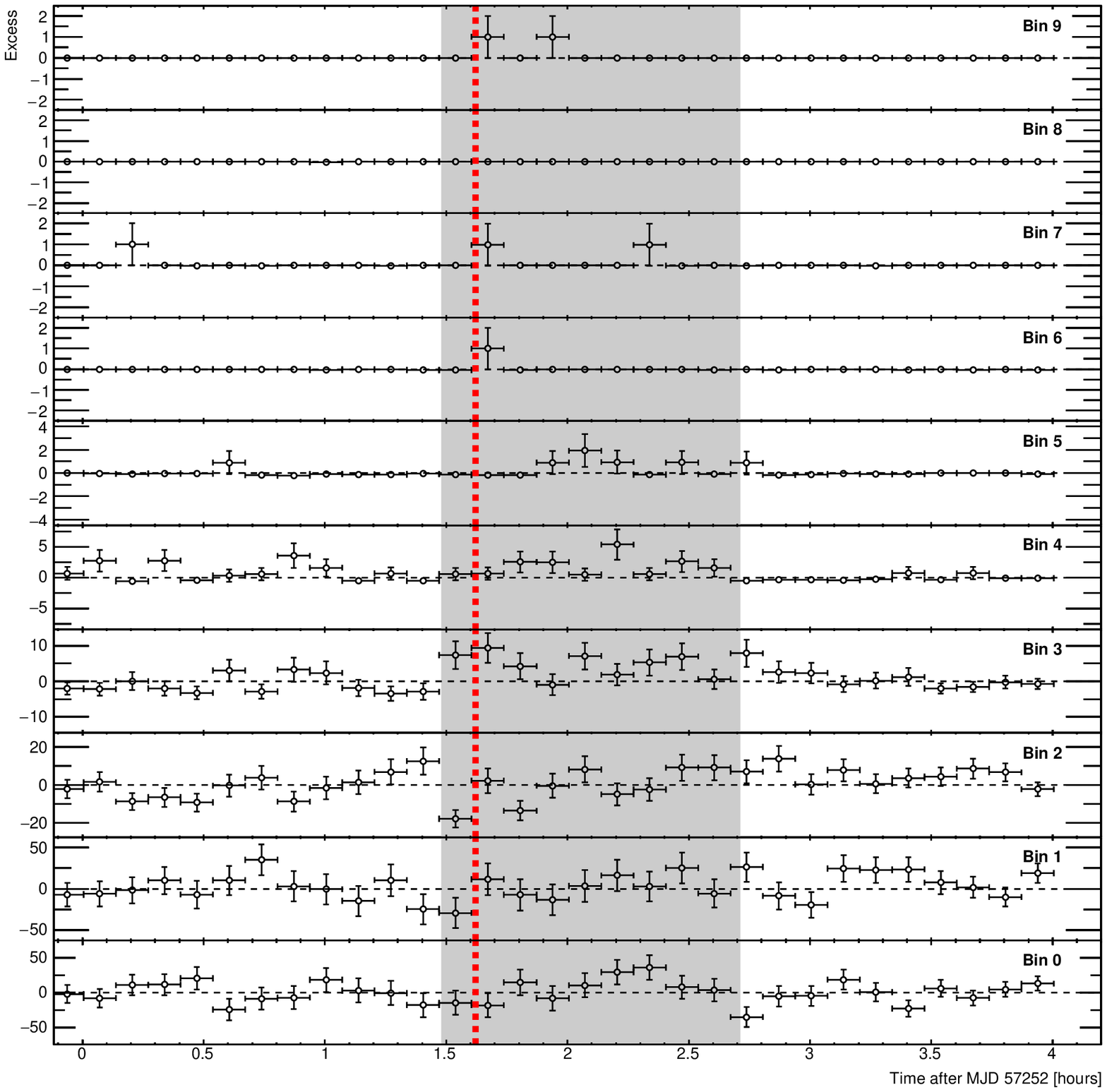}
\caption{
Excess counts above background for times around the flare from Mrk 501 on MJD 57252, in bins of duration 8 minutes for clarity.
From bottom to top, the panels show the excess in analysis bins 0 through 9.
The vertical red dashed line marks the flare start time identified by the search algorithm.
The gray shaded area shows the duration of the flare as determined by the Bayesian block algorithm of \citet{Scargle:2013dm}.
}
\label{sec:verify:fig:mrk501-flare57252-excess}
\wfigend

Figure~\ref{sec:verify:fig:mrk501-flare57252-excess} shows the light curve for the flare in terms of excess counts above background for each of the ten analysis bins.
The start of the flare as determined by the flare monitor search algorithm appears as a dashed red vertical line in the figure.
Analysis bins 3, 4, 5, 7, and 9 dominate the contributions to the test statistic $D$ from Equations~\ref{sec:method:eqn:bin-ts-trigger} and \ref{sec:method:eqn:full-ts-trigger}, together accounting for 96\% of the total.
This assessment of $D$ uses all observations in the buffer, including data from the previous transit of the source.
The inclusion of these data produces a more precise estimate of the background event rate than if we used only data collected over a shorter period.
This procedure improves the algorithm's determination of the importance of the higher analysis bins where the events are sparse.
Our confidence in this event being a real flare increases because the contributions to $D$ come from several analysis bins and are not dominated by a single bin.

As a consistency check, we also run the Bayesian block algorithm of \citet{Scargle:2013dm}, which among other applications is commonly used to identify significantly distinct states in astrophysical light curves \citep[e.g.][]{Barriere:2014dr,Aartsen:2015jx,Ackermann:2016gt}.
The advantages of this algorithm include insensitivity to gaps in the data taking and, for a given set of $N$ observations, the capacity to identify the optimum partition of the data in $N^2$ steps.
Because of this scaling, it is possible to run the Bayesian block algorithm over the few thousand 2-minute points that fit into several days, but it is computationally prohibitive to run the algorithm over the complete data set of $\sim$1.5 years with 2-minute time resolution.

Following Section 3 of \citet{Scargle:2013dm}, we use the full likelihood ratios constructed by Equations~\ref{sec:method:eqn:bin-ts} and \ref{sec:method:eqn:full-ts} to identify regions of the light curve where the flux from the source is well represented by a constant value.
Our threshold parameter $\gamma$ from Equation~\ref{sec:method:eqn:trigger-condition} naturally becomes the prior parameter required by the Bayesian block algorithm.
For a given set of observations, $\gamma$ controls the probability that the Bayesian block algorithm will produce a single false trigger.

We select four separate time windows centered on the identified start point of the flare, with durations of 2, 4, 6, and 8 days.
For each separate time window, we run the Bayesian block algorithm with a value of $\gamma$ such that the probability for a false trigger is 5\%.
We choose a relatively high probability because our goal here is to identify the duration of a flare that has already been detected by an independent method, not to quantify the significance of our observation.

All of the windows produce a strong detection of a flare beginning 88~$\pm$~2 minutes after the start of MJD 57252 and ending 74~$\pm$~2 minutes later.
We also run the Bayesian block algorithm with a false trigger rate of 1\%.
In this more restrictive search, only the 6-day and 8-day time windows detect the flare.
It is likely that the shorter time windows do not contain sufficient information in the region before and after the flare to overcome the stricter trigger condition.

The consistency between the various windows gives us confidence that the trigger is not due to a peculiarity of the data on MJD 57252, but is a real flare that does not depend on where the edges of the data selection occur.
The gray band in Figure~\ref{sec:verify:fig:mrk501-flare57252-excess} shows the block that is selected by the Bayesian block algorithm for the 8-day time window.
The start of this region lines up well with the starting point identified by the faster but more limited algorithm used by the flare monitor.

\figstart
\includegraphics[width=\figurewidth]{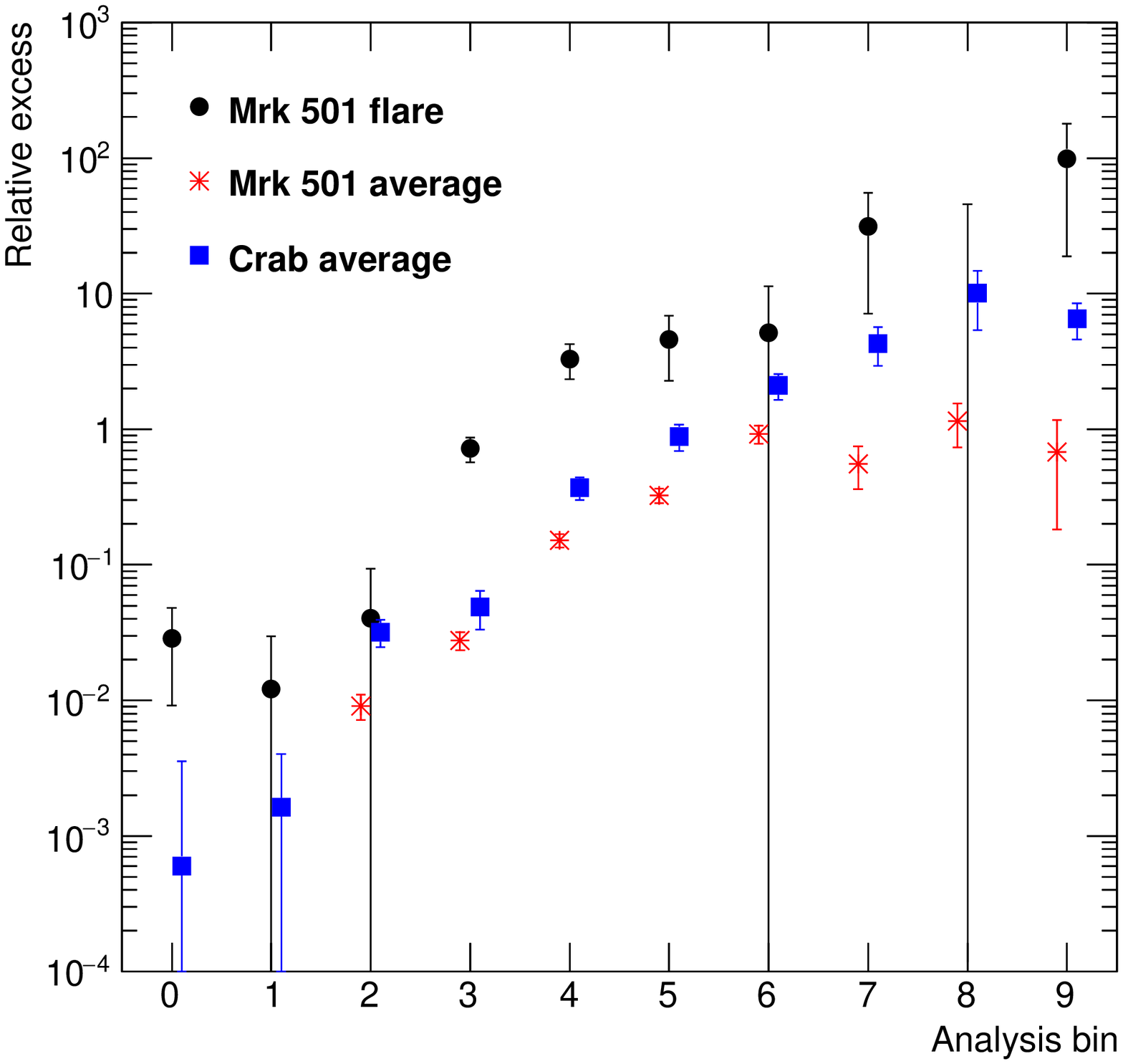}
\caption{
Relative excess (black circles) as a function of analysis bin for the Mrk 501 flare on MJD 57252.
For reference, we include the long-term relative excess from the entire data set for the Crab Nebula (blue squares) and for Mrk 501 (red asterisks).
These points are offset slightly from the analysis bin for clarity.
}
\label{sec:verify:fig:mrk501-flare57252-relative-excess}
\figend

Figure~\ref{sec:verify:fig:mrk501-flare57252-relative-excess} shows the relative excess as a function of analysis bin for the data collected from Mrk 501 during the flare identified by the Bayesian block algorithm.
The long-term relative excess from the Crab Nebula taken over the entire data set is shown for comparison, and the long-term relative excess from Mrk 501 over the entire data set also appears in the figure.
These long-term relative excesses have been computed from a weighted average that matches the zenith angle distribution to the observed distribution from Mrk 501 during the flare.
Furthermore, the background used to compute the relative excess during the flare has been derived from zenith-matched data collected 2.5 days before and after the flare.
This procedure was necessary to insure that a sufficient number of background events were collected to produce an accurate determination of the relative excess in the higher analysis bins.
In addition to deriving the background counts from 2.5 days before and after the flare, we have also checked the results using 10 days before and after the flare and the full data set.
We have also derived the background from the time of the flare alone.
We find good agreement between the relative excesses for all analysis bins in each case.

As discussed in Section~\ref{sec:hawc-operation}, the analysis bins do not correspond directly to energy, but rather to the amount of information available about the events.
Deriving a spectrum requires the detailed analysis techniques presented in \citet{HAWC-Crab-paper} that include the HAWC detector response and account for the energy overlap and zenith angle response of the analysis bins.
Consequently, the temptation to interpret the data in Figure~\ref{sec:verify:fig:mrk501-flare57252-relative-excess} in terms of energy spectra should be avoided, as such interpretations will not produce reliable results.
Rather, we include Figure~\ref{sec:verify:fig:mrk501-flare57252-relative-excess} to demonstrate that the detected flare is substantially stronger than both the average Mrk 501 flux and the Crab Nebula flux.

\figstart
\includegraphics[width=\figurewidth]{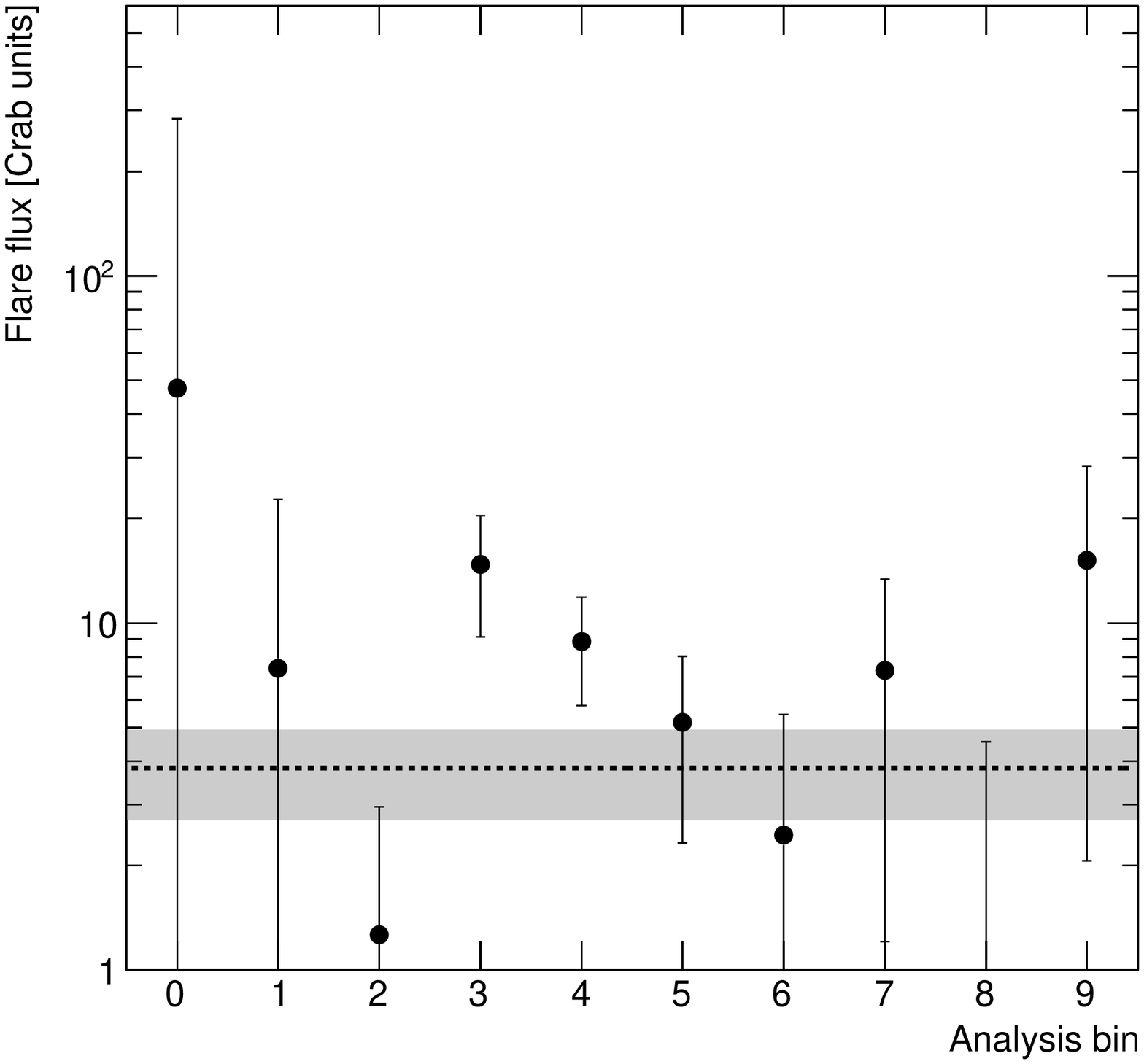}
\caption{
Flare flux scaled to the Crab Nebula flux as a function of analysis bin for the Mrk 501 flare on MJD 57252.
The dashed line and shaded error band result from a fit to a constant value across all analysis bins.
}
\label{sec:verify:fig:mrk501-flare57252-crab-units}
\figend

Since the flare monitor is sensitive to multi-Crab level flares lasting on the scale of hours, we expect the detected flare to surpass both the flux from the Crab and the long-term average flux from Mrk 501.
We obtain a rough estimate of the flux of the flare in Crab units by fitting a constant value to the ratio of the relative excess from Mrk 501 during the flare to that from the Crab Nebula over the full data set.
This estimate is only approximate because the analysis bins overlap substantially in energy and the spectrum of the flare may differ from the Crab spectrum.
Nevertheless, the errors on the relative excess ratio are large enough to admit a good fit with a constant value, as shown in Figure~\ref{sec:verify:fig:mrk501-flare57252-crab-units}.
The flare flux according to the fit is 3.8~$\pm$~1.1 Crab units, in accord with expectations based on the sensitivity of the method presented in Figure~\ref{sec:sensitivity:fig:base-sensitivity}.

\subsection{Summary of Mrk 421 and Mrk 501 Triggers}\label{sec:verify:sub:trigger-summary}

\begin{deluxetable*}{cccccccc}
\tablecaption{High-Confidence Markarian Triggers Identified by the Archival Search\label{sec:verify:tab:mrk-flares}}
\tablehead{
\colhead{} & \colhead{False Alarm} & \colhead{} & \colhead{Trigger} & \colhead{Time to} & \colhead{Search} & \colhead{Bayesian Block} & \colhead {Approximate} \\
\colhead{Event} & \colhead{Rate [$\mathrm{century}^{-1}$]} & \colhead{Target} & \colhead{MJD} & \colhead{Detection [min.]} & \colhead{Threshold} & \colhead{Duration [min.]} & \colhead{Flux [Crab units]}
}
\startdata
1 & 0.68 & Mrk 501 & 57252.068 & 34 & 1\% & 74 & 3.8~$\pm$~1.1 \\
2 & 2.1 & Mrk 421 & 57020.357 & 112 & 5\% & 158 & 3.9~$\pm$~0.8 \\
3 & 2.7 & Mrk 501 & 57032.614 & 80 & 5\% & 148 & 2.1~$\pm$~0.6 \\
\enddata
\tablecomments{\editadded{The false alarm rate calculation ignores the presence of other sources and therefore represents half of the total false alarm rate for the Mrk 421 \& 501 source category. Time to detection refers to the amount of time between the start of the flare and the time at which it was identified. The search threshold indicates whether the event detected in the Bayesian block analysis with an 8-day window appears at the $1\%$ false alarm level or only at the $5\%$ false alarm level. The Bayesian block duration is the duration of the flare as identified in that analysis, and the approximate flux is computed during this duration.}}
\end{deluxetable*}

In the data set considered in this work, we expect 3.8 events surpassing our trigger threshold in the Mrk 421 \& Mrk 501 target class.
We observe a total of six events: four from Mrk 421 and two from Mrk 501.
Three of these events have large equivalent false alarm rates and appear consistent with background fluctuations.
Here, we consider the three other events, which all have false alarm rates much lower than we would expect from background alone.
We designate these as {\it high-confidence} triggers from the Markarian source class.

For each of the three high-confidence triggers, we run the Bayesian block algorithm with an 8-day window in order to estimate the flux following the methods of Section~\ref{sec:verify:sub:flare57252}.
If the flare is not identified in the Bayesian block algorithm with a 1\% false trigger rate, we run the algorithm with a more permissive false trigger rate of 5\%.
Since our goal is to identify the properties of events that have already been detected by an independent algorithm and not to quantify the significance of our detections, we are not concerned with statistical trials in this case.

Table~\ref{sec:verify:tab:mrk-flares} summarizes the properties of the three high-confidence events.
The columns in Table~\ref{sec:verify:tab:mrk-flares} show: (1) a convenient reference number for the event, (2) the equivalent false alarm rate for the event at its most significant detection, (3) whether the event is associated with Mrk 421 or Mrk 501, (4) the MJD of the trigger time, (5) the time to detection for the event, (6) whether the event is identified in the Bayesian block search run with an 8-day window at the 1\% false alarm or 5\% false alarm level, (7) the duration of the flare as identified by the Bayesian block search, and (8) the approximate flux of the flare as determined by the methods of Section~\ref{sec:verify:sub:flare57252}.

Event 1 in Table~\ref{sec:verify:tab:mrk-flares} is the Mrk 501 flare from Section~\ref{sec:verify:sub:flare57252} and has the lowest equivalent false alarm rate.
Events 2 and 3 are associated with Mrk 421 and Mrk 501 respectively.
These events have false alarm rates in the range of 2 to 3 events per century, and their estimated durations and fluxes in Crab units are in agreement with the expectations from Section~\ref{sec:sensitivity}.
Furthermore, their times to detection are smaller than their durations.
Thus all three of these events would have been detected by the real-time flare monitor sufficiently rapidly to issue alerts for follow-up observations, had the flare monitor been operational at the time.

The HAWC collaboration recently reported on a separate search of Mrk 421 and Mrk 501 for flares using the standard HAWC analysis with a base time scale of 1 transit \citep{HAWC-LC-paper}.
This search is sensitive to weaker flares but lacks the rapid response and flexible time scale search of the real-time flare monitor.
In \citet{HAWC-LC-paper}, several periods of significant enhanced activity appear in both sources.
The three high-confidence events identified by the present work all correspond to regions of elevated flux according to the standard HAWC analysis, providing further support for the detections presented in this work.
It is plausible that the real-time flare monitor, which is dedicated to finding short, strong flares, fails to detect the other flares identified by the single-transit light curves because they are weaker in flux.
Flares that rise slowly over the course of several days would also elude the present analysis due to its focus on short time scales.
Further assessment of the consistency of the two methods is ongoing.

\editadded{
Event 1 in Table~\ref{sec:verify:tab:mrk-flares} occurred during quasi-simultaneous observations from the FACT experiment~\citep{Anderhub:2013hb}.
The publicly available light curves from FACT\footnote{\editadded{See the quick-look analysis at \url{http://fact-project.org/monitoring}.}} show a clear increase in the excess rate from Mrk 501 on the same night as event 1.
This excess is significantly larger than the long-term FACT excess measured from Mrk 501, as well as the FACT excess measured on the surrounding nights.
The FACT observations took place between MJD 57251.87 and MJD 57251.96, approximately 2.6 hours before the flare detected by HAWC at MJD 57252.068.
Owing to the short delay between the flare detections by the two instruments, it is likely that the increased activity evident in the FACT observations is related to the subsequent outburst detected by HAWC a few hours later.
}

\editadded{
Near the time of event 2, which is associated with Mrk 421, FACT observations are available on MJD 57015 and 57021, with the latter spanning the range from MJD 57021.14 to 57021.29.
These observations occurred approximately 17 hours after the end of event 2 as identified by HAWC.
Nevertheless, the available FACT monitoring data show that Mrk 421 was in a relatively high state compared to its long-term average during these dates, in agreement with the daily HAWC observations reported by \cite{HAWC-LC-paper}.
The HAWC event detected on MJD 57020 may be related to this high state.
}

\editadded{
Event 3 occurred when the source was not visible to FACT during nighttime observations.
As a result, FACT monitoring data are unavailable for event 3.
}

\section{Flare Monitor Outlook}\label{sec:outlook}

The HAWC real-time flare monitor has been fully operational since 2017 January, when it began sending alerts to the VERITAS experiment.
Alerts to the Fermi-LAT began in 2017 March, and additional instruments are currently being added.
Ultimately, we aim to provide public alerts via the Gamma-ray Coordinates Network~\citep{Barthelmy:dp}.

This work has explored the capability of the HAWC real-time flare monitor to detect VHE flares at the level of several times the flux from the Crab Nebula in under 1 hour.
We have demonstrated this capability via the detection in archival HAWC data of three high-confidence flares from Mrk 421 and Mrk 501.
These observations are consistent with fluxes several times that from the Crab Nebula and with durations ranging from one to a few hours.
We have also shown that the false alarm rate of the method matches the expectation from simulation, allowing us confidently to assign an equivalent false alarm rate for each detection.

The HAWC real-time flare monitor will provide a unique opportunity to study an unbiased collection of extreme VHE flares in order to unravel further the mysteries surrounding blazars, their VHE emission, and their jets.
Of particular interest are the nature and location of the dissipation region, where the flares are produced.
Understanding the particle populations and acceleration mechanisms that produce the flares, investigating the possible correlation of X-ray and gamma-ray fluxes during the flares, and determining the causes of apparent orphan VHE flares all proceed from the detection of these events.
By helping to identify the most extreme flares and the shortest time scales on which they occur, the HAWC real-time flare monitor will facilitate coordinated observations seeking new clues for addressing these and other important topics of interest.

\acknowledgements

We acknowledge the support from:
the US National Science Foundation (NSF);
the US Department of Energy Office of High-Energy Physics;
the Laboratory Directed Research and Development (LDRD) program of Los Alamos National Laboratory;
Consejo Nacional de Ciencia y Tecnolog{\'\i}a (CONACyT), M{\'e}xico (grants 271051, 232656, 260378, 179588, 239762, 254964, 271737, 258865, 243290, 132197), Laboratorio Nacional HAWC de rayos gamma;
L'OREAL Fellowship for Women in Science 2014;
Red HAWC, M{\'e}xico;
DGAPA-UNAM (grants IG100317, IN111315, IN111716-3, IA102715, 109916, IA102917);
VIEP-BUAP;
PIFI 2012, 2013, PROFOCIE 2014, 2015;
the University of Wisconsin Alumni Research Foundation;
the Institute of Geophysics, Planetary Physics, and Signatures at Los Alamos National Laboratory;
Polish Science Centre grant DEC-2014/13/B/ST9/945;
Coordinaci{\'o}n de la Investigaci{\'o}n Cient{\'\i}fica de la Universidad Michoacana.
Thanks to Luciano D{\'\i}az and Eduardo Murrieta for technical support.

\facility{HAWC}

\bibliography{manuscript}

\end{document}